\documentclass[preprint]{aastex}
\shorttitle{ Validation of {\it Kepler}'s Multiple Planet Candidates. II.  }
\shortauthors{ Lissauer et al.}
\usepackage{natbib}

\newcommand{\ikt}{{\it Kepler}}
\newcommand{\ik}{{\it Kepler~}}

\begin{document}
\slugcomment{Astrophysical Journal, in press }
\title{ Validation of \ikt's Multiple Planet Candidates. II: Refined Statistical Framework and Descriptions of Systems of Special Interest}

\author{Jack J. Lissauer\altaffilmark{1}, Geoffrey W. Marcy\altaffilmark{2},  Stephen T. Bryson\altaffilmark{1}, Jason F. Rowe\altaffilmark{1,3}, Daniel Jontof-Hutter\altaffilmark{1}, Eric Agol\altaffilmark{4},   William J. Borucki\altaffilmark{1}, Joshua A. Carter\altaffilmark{5}, Eric B. Ford\altaffilmark{6,7}, Ronald L. Gilliland\altaffilmark{6,7}, Rea Kolbl\altaffilmark{2}, Kimberly M. Star\altaffilmark{6,7}, 
 Jason H. Steffen\altaffilmark{8}, Guillermo Torres\altaffilmark{5}
}

\email{ Jack.Lissauer@nasa.gov }
\altaffiltext{1}{NASA Ames Research Center, Moffett Field, CA 94035, USA}
\altaffiltext{2}{Astronomy Department, University of California, Berkeley, CA 94720, USA}
\altaffiltext{3}{SETI Institute, Mountain View, CA 94043, USA}
\altaffiltext{4}{Department of Astronomy, Box 351580, University of Washington, Seattle, WA 98195, USA}
\altaffiltext{5}{Harvard-Smithsonian Center for Astrophysics, 60 Garden Street, Cambridge, MA 02138, USA}
\altaffiltext{6}{Department of Astronomy and Astrophysics, 525 Davey Laboratory, The Pennsylvania State
University, University Park, PA 16802, USA}
\altaffiltext{7}{Center for Exoplanets and Habitable Worlds, 525 Davey Laboratory,
The Pennsylvania State University, University Park, PA, 16802, USA}
\altaffiltext{8}{Department of Physics $\&$ Astronomy/CIERA, Northwestern University, 2145 Sheridan Road, Evanston, IL 60208}
\begin{abstract}
We extend the statistical analysis of Lissauer et al.~(2012, ApJ 750, 112), which demonstrates that the overwhelming majority of \ik candidate multiple transiting systems (multis) represent true transiting planets, and  develop therefrom a procedure to validate large numbers of planet candidates in multis as {\it bona fide} exoplanets.  We show that this statistical framework correctly estimates the abundance of false positives already identified around \ik targets with multiple sets of transit-like signatures based on their abundance around targets with single sets of transit-like signatures. We estimate the number of multis that represent split systems of one or more planets orbiting each component of a binary star system. We use the high reliability rate for multis to validate more than one dozen particularly interesting multi-planet systems 
herein.  Hundreds of additional multi-planet systems 
are validated in a companion paper by Rowe et al. (2014, ApJ ).  We note that few very short period ($P < 1.6$ days) planets orbit within multiple transiting planet systems and discuss possible reasons for their absence. There  also  appears to be a shortage of planets with periods exceeding a few months in multis. 
\end{abstract}
 
\clearpage
\section{ Introduction }

More than 40$\%$ of the \ik planet candidates announced by \cite{Borucki:2011}, \cite{Batalha:2013} and \cite{Burke:2013} are associated with 
targets that have more than one candidate planet.  
Accounting for candidates on each one's individual merit, without considering whether the same target has any additional candidates, \cite{Morton:2011} estimated the fidelity of \ikt's planet 
candidates (fraction of the candidates expected to be actual planets) to be above $90\%$. Incorporating the corrections presented in \cite{Santerne:2013}, \cite{Fressin:2013} derive a slightly larger fraction of false positives (FPs), although they classify planets orbiting stars other than the \ik target to be FPs. If one doesn't consider such planets to be FPs, then their estimate of the number of FPs is reduced by more than a factor of two, leading to a planet fidelity rate of $\sim 95\%$. In this work, we do not classify planets orbiting a star that is bound to the \ik target to be FPs, and among the candidates that we validate as planets, we expect very few, if any, planets orbiting  stars unrelated to the \ik target. The galactic latitude distribution of \ik targets with both single and multiple planet candidates tracks that of \ik targets, but is quite different from that of false positives produced by chance-alignment blends with faint eclipsing binary stars (Appendix A), increasing our confidence in the \ik planet candidate sample. 

\ik has found far more multiple planet candidate systems (multis) than would be the case if candidates were randomly 
distributed among target stars \citep{Lissauer:2011b, Latham:2011}. \cite{Lissauer:2012}, henceforth Paper I, presented a statistical analysis that combined the large numbers of multis observed by \ik  (as listed in \citealt{Borucki:2011}) together with the assumption that false positives are nearly 
randomly distributed among \ik targets to demonstrate that  the fidelity of \ik
multiple planet candidates is far higher than that for singles.  We expand upon the statistical  analysis of Paper I herein and show that it correctly accounts for the number of FPs that have been identified in multis.  The high reliability of the remaining sample forms the foundation of a framework that we develop to validate a large number of \ikt 's multiple planet candidates as  true planets. We present the techniques that we use to search for companions of planet-hosting stars, and then validate and discuss  multi-planet systems of high multiplicity and/or special dynamical interest.  A companion paper by \cite{Rowe:2014}, henceforth Paper III, 
describes the light curve studies and astrometric measurements of transit locations using pixel-level \ik data to reveal the position of the center of light (centroid) of the \ik target star both during transits and outside of transit, along with the classification of planet-hosting stars. Paper III validates hundreds of new multiple planet systems.

 In Section 2, we refine the statistical analysis presented within Paper I. We discuss the ensemble of candidate planetary systems considered for validation, which is based upon multis identified using the first two years of \ik data, in Section 3. We combine the equations of Section 2 with the candidate pool presented in Section 3 to yield numerical estimates of the number of FPs present in multis in Section 4. Section 4.2 describes tests of the statistical framework for assessing the expected number of FPs in multis using the distributions of \emph{identified} FPs and remaining planet candidates. In Section 5, we consider and classify the various sources that can produce multiple, periodic, transit-like signatures in the light curves of \ik target stars, with the principal sources being transiting planets and eclipsing binary stars. We provide estimates of the fraction of multis that represent blended planetary systems (including planets orbiting each of the components of a binary star system)  in Section 6. The criteria that we use to validate \ik planet candidates as {\it bona fide} planets are presented in Section 7. In Section 8, we present  our search for stellar multiplicity in these systems. The three planets within binary star system Kepler-132 = KOI-284, with one of the stars possessing two transiting planets and the other star being orbited by a single transiting planet, as well as the five planets on S-type (circumstellar) orbits in the Kepler-296 = KOI-1422 binary system, are validated and discussed in Section 9. Section 10 summarizes the properties of other \ik systems with five or more transiting planet candidates, beginning with the seven-planet Kepler-90 = KOI-351 system in Subsection 10.1.  The  Kepler-223 = KOI-730 planetary system with four planets very close to a ``chain'' of mean motion resonances is presented in Section 11.   
 We conclude the main text by summarizing our principal results in Section 12.
 
 Appendix A compares the distributions in galactic latitude of \ik targets, planet candidates and false positives. Appendix B discusses the differing period distributions of various classes of transit-like signatures that have been identified in \ik data and the implications of these differences for planet validation and planetary system architecture. Computing false positive rates by comparing overall \ik planet candidate abundances to abundances of eclipsing binary stars severely underestimates the reliability of multis but overestimates the reliability of single planet candidates. The ramifications of the large abundance and high reliability of multis on estimates of the reliability of \ik singles  are addressed in Appendix C.  Appendix D presents the  dynamical simulations that we performed to assess stability of systems with tightly-packed planet candidates as well as those to determine which of the candidate planets that are large enough to be brown dwarfs can be demonstrated to be planets by limits on transit timing variations (TTVs) induced in neighboring planets.

\section{ Statistical Framework }
We begin our quantitative analysis by deriving formulas for the expected numbers of false positives in multis assuming that FPs are randomly distributed among \ik targets.  Our results are expressed in terms of the numbers of targets, planet candidates and multis, the expected FP rate for the single planet candidates, etc.  These formulas are evaluated numerically in Section 4 (see Tables  \ref{tab:fpest} and  \ref{tab:stattest}).  We use the following notation: 
 
\noindent $n_t \equiv \#$ of targets from which the sample is drawn   

\noindent $n_i \equiv \#$ of targets with exactly $i$ candidates, where $i$ is a positive integer; e.g., $n_1 \equiv \#$ of targets with a single planet candidate

\noindent $n_k \equiv \#$  of \ik targets with one or more candidates $= \sum_{i=1}^\infty n_i$

\noindent $n_m \equiv \#$ of targets with two or more candidates $ = \sum_{i=2}^\infty n_i$, i.e., the number of candidate multiple transiting planet systems,  so $n_k = n_m +  n_1$
 
\noindent  $n_c \equiv \#$ of planet candidates $= \sum_{i=1}^\infty i n_i$
 
\noindent $n_p \equiv \#$ of actual planets among the candidates 

\noindent $n_{p,1} \equiv \#$ of actual planets among the single candidates 

\noindent $n_{\rm f} \equiv \#$ of false positives  among the candidates,  so $n_c = n_p +  n_{\rm f}$

\noindent $n_{\rm fm} \equiv$  the number of false positive planet candidates present among the multis

\noindent ${\cal P} \equiv n_p/n_c$ is the fidelity of the sample (fraction of candidates that are planets) 

\noindent ${\cal P}_1 \equiv n_{p,1}/n_1 \equiv$ the fidelity  of the sample of single planet candidates

\noindent $\lambda  \equiv$ the average number of false positives per target $= n_{\rm f}/n_t = (n_c - n_p)/n_t = ((1-{\cal P}_1)n_1 + n_{\rm fm})/n_t$.

As in Paper I, we make the following two assumptions for  our statistical estimates of the number of FPs in multis:

\noindent 1) FPs are {\it randomly distributed} among the targets; 

\noindent 2) there is {\it no correlation} between the probability of a target to host one or more detectable 
planets and to display FPs.

\noindent See Sections 5 and 6 of this paper and Section 2.2 of Paper I for  discussions of these assumptions.
Paper I used these two assumptions, together with observed values of $n_c$, $n_m$ and $n_t$ based on the KOI (\ik object of interest) planet candidate catalog in \cite{Borucki:2011} with the minor modifications presented in \cite{Lissauer:2011b} and an 
assumed lower bound on ${\cal P}$, to estimate lower bounds on the fidelity of candidates in various 
classes of candidate multi-planet systems.  Our methodology is similar, except that we assume a lower bound on ${\cal P}_1$ rather than on ${\cal P}$ and we use a more recent, larger, more uniform and better vetted set of planet candidates for our studies.  

Assumption (1) implies 
that the expected number of targets with $j$ false positives, $E(j)$, is given by a Poisson distribution of $n_t$  
members whose mean is given by $\lambda n_t$.  The expected number of targets with $j$ false positives is given by the following formula: 

\begin{equation}
 E(j) = \frac{\lambda^j e^{-\lambda }}{j!} n_t  \approx  \frac{\lambda^j}{j!} n_t ,
\end{equation}
 
\noindent where the approximation in Equation (1) is valid for $\lambda \ll 1$. This is clearly a good approximation  because the number of planet candidates is less than two percent as large as the number of targets  and the number of false positives  by definition cannot exceed the number of planet candidates ($n_{\rm f}/n_t \le n_c/n_t < 0.02$). 

We  compute below the expected number of \ik multis in which  at least one candidate is a  FP.  In some cases, we make approximations that yield conservative estimates of the expected numbers 
of true planets (i.e., overestimate the expected number of FPs). 
Our results, presented in Equations (\ref{eq:2fp}) -- (\ref{eq:2p2fp}), are given as general formulae\footnote{In principle, these estimates should be expressed as inequalities, but since the corrections are very small compared to the uncertainties in the numerical estimates of the abundance of FPs in singles that are required as input, we simply state them as equalities.}; numerical values for a specified subset of \ik candidates and an assumed   
fraction  of the singles that are true planets, ${\cal P}_1$, are provided in Section 4.1 (see Table \ref{tab:fpest}). Section 4.2 demonstrates that these formulas correctly estimate the number of identified FPs in multis given the identified FP rate for singles (see Table \ref{tab:stattest}).

Equation (1) yields estimates of the number of targets with two or three false positives:

\begin{equation}
\label{eq:2fp}
2 ~{\rm FPs:}~ E(2) = \frac{((1-{\cal P}_1)n_1 + n_{\rm fm})^2}{2n_t} ;
\end{equation}
\begin{equation}
\label{eq:3fp}
 3 ~{\rm FPs:}~ E(3) = \frac{((1-{\cal P}_1)n_1 +n_{\rm fm})^3}{6n_t^2} ,
\end{equation}

\noindent  where $n_{\rm fm}$ provides an upper bound on the number of targets that are classified as multis but actually contain a single planet and at least one  false positive. The expected number of targets with four or more false positives is extremely small and can be 
neglected for our purposes. Note that the expected number of targets with no planets and a single FP under the same assumptions is 
$$0~{\rm planets}~ + 1~{\rm FP:}~ E(1) = (1-{\cal P}_1)n_1 .$$

The assumed lack of correlation between the propensity of a target to have false positives and true 
planets implies of the probability of a given target hosting both a planet and one or more false positives 
is equal to the product of these individual probabilities.  The probability
of a target having a single detected transiting planet is bounded from above by $({\cal P}_1n_1+n_{\rm fm})/n_t$, and the probability of it showing a false positive
is likewise bounded by $\lambda$.  Thus, the estimated numbers of targets with at least one planet as well as one or more false positives 
are given by:
  
\begin{equation}
\label{eq:1p1fp}
1 ~{\rm planet}~ + 1~{\rm FP:}~ \frac{n_1{\cal P}_1+n_{\rm fm}}{n_t}\times \frac{n_k(1-{\cal P}_1)}{n_t} \times n_t = \frac{(n_1{\cal P}_1+n_{\rm fm})(1-{\cal P}_1)n_k}{n_t};
\end{equation}
\begin{equation}
\label{eq:1p2fp}
1 ~{\rm planet}~ + 2~{\rm FPs:}~ \frac{n_1{\cal P}_1+n_{\rm fm}}{n_t}\times \frac{((1-{\cal P}_1)n_1 + n_{\rm fm})^2}{2n_t}  = \frac{(n_1{\cal P}_1+n_{\rm fm})((1-{\cal P}_1)n_1 + n_{\rm fm})^2}{2n_t^2}.
\end{equation}

\noindent The expected number of targets with both a planet and three or more false positives is very 
small and can be neglected for our purposes.  We note that no targets have yet been found to host more than two FPs (Table \ref{tab:stattest}).

To derive an estimate of the expected number of targets with multiple true planets as well as at least one 
FP, replace the first term in Equations (\ref{eq:1p1fp}) and (\ref{eq:1p2fp}), $(n_1{\cal P}_1+n_{\rm fm})/n_t$, with a term representing the probability of a given target having a multi-planet 
system, $n_m/n_t$. (Our use of $n_m/n_t$  leads to a slight over-estimate of the number of false positives among the multi-candidate systems,
as it assumes all such systems contain at least two real transiting planets.)
 \begin{equation}
\label{eq:2p1fp}
{\rm 2~or~more~planets}~ + 1~{\rm FP:}~ \frac{n_m}{n_t}\times \frac{n_k(1-{\cal P}_1)}{n_t} \times n_t  = \frac{n_mn_k(1-{\cal P}_1)}{n_t};
\end{equation}
\begin{equation}
\label{eq:2p2fp}
{\rm 2~or~more~planets}~ + 2~{\rm FPs:}~ \frac{n_m}{n_t}\times \frac{((1-{\cal P}_1)n_1 + n_{\rm fm})^2}{2n_t}  =\frac{n_m((1-{\cal P}_1)n_1 + n_{\rm fm})^2}{2n_t^2}.
\end{equation}

\noindent The expected number of targets with both multiple planets and three or more false 
positives is extremely small and can be neglected for our purposes. 

\section{Candidate Multiple Planet Systems Considered for Validation}

The total planet candidate pool analyzed for our statistical study is based on the catalog of \ik planet candidates identified using the first two years of \ik data (Q1-Q8)\footnote{ The 
\ik spacecraft rotated four times per $\sim$372 day orbit to keep its sunshade and solar panels oriented properly.  Targets were imaged on different parts of the 
focal plane during different orientations.  The data are grouped according to the ``quarter'' year during 
which observations were made, with the first group labeled Q1 and subsequent quarters  
numbered sequentially: Q2, Q3, ....}.  We consider for validation by multiplicity only systems that contain at least two candidates that meet all of the following criteria.  These requirements are applied uniformly to single and multiple planet system candidates for our statistical study.

($i$) It was identified in the Q1 -- Q8 automated transit search and/or listed in \cite{Batalha:2013}.

($ii$) It is not labeled as a known FP.

($iii$) The pattern of transits has a cumulative signal-to-noise ratio S/N $>$ 10 (through Q14).

($iv$) A minimum of three transits have been observed (through Q15).

($v$) The orbital period $P > 1.6$ days.

($vi$) The transit light curve is U-shaped (rather than V-shaped). We define a transit to be U-shaped if the best fit to the transit light curve implies that the entire planet is in front of the stellar disk at mid-transit  with a confidence exceeding one standard deviation in estimated impact parameter, i.e., 
 \begin{equation}
b + b_\sigma + \frac{R_p}{R_\star} < 1, 
 \end{equation}
where $b$ is the impact parameter, $b_\sigma$ is its + 1 standard deviation (uncertainty), and $R_p$ and $R_\star$ are the radii of the planet and the star, respectively.

\begin{figure}
\includegraphics[scale=0.45]{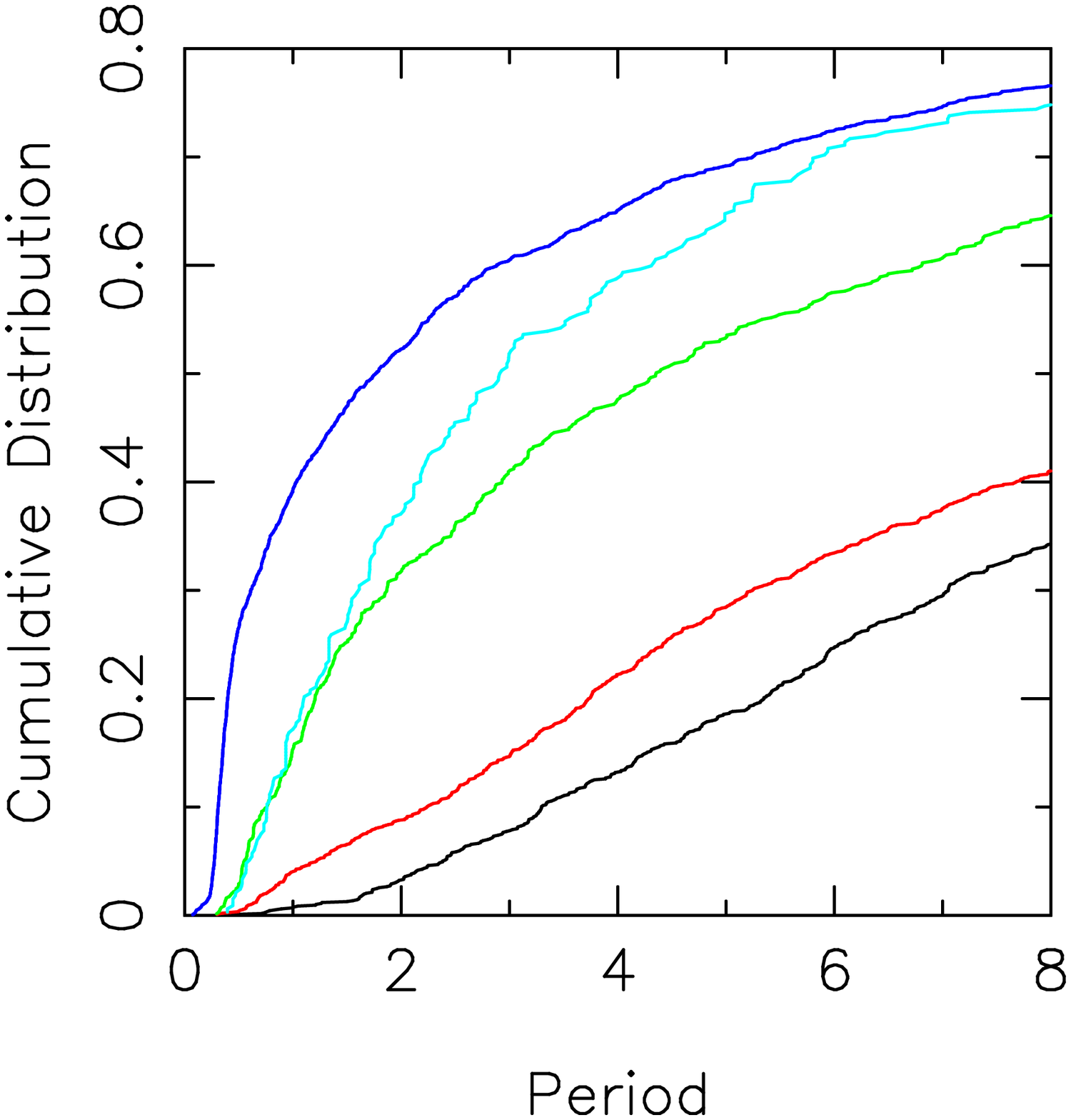}
\includegraphics[scale=0.45]{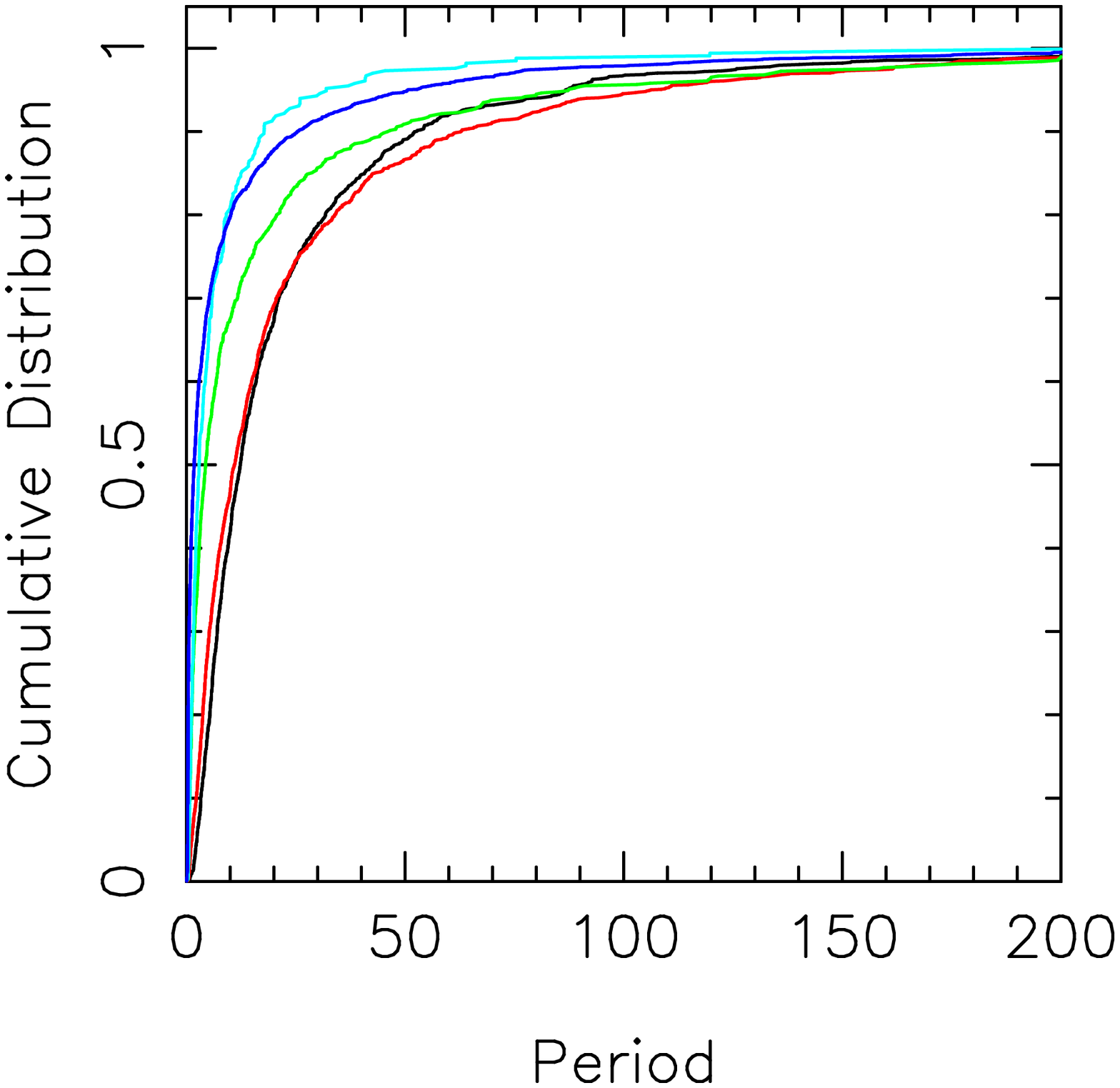}
\caption{The normalized cumulative period distributions of various transit-like patterns observed by \ikt. The five curves represent distinct (non-overlapping) populations. The red curve shows single planet candidates and the black curve planet candidates in multis. The other curves display various types of signatures of EBs as follows: The blue curve displays the distribution of detached eclipsing binaries that were identified as such rapidly and thus were not given KOI numbers, the turquoise curve shows FPs that were produced by instrumental effects distributing light from bright targets and identified by matches in period and epoch \citep{Coughlin:2013}, the green curve indicates other false positives. Each curve has been normalized by the total number of objects of its class that have been detected. The normalization includes all orbital periods, but the plots only show the cumulative distributions up to 8 days in the left panel and up to 200 days in the right panel (an analogous plot up to 40 days is shown in Paper III).  To remove spurious detections of noise, we impose criteria ($iii$; SNR $>$ 10) and ($iv$; minimum of 3 events observed) of Section 3; criterion ($i$; identified in Q1-Q8 data) is also required for all populations apart from the EBs.
}
\label{fig:perdist}
\end{figure}

Criterion ($i$) is employed to minimize bias in identification of singles relative to multis.  Criterion ($iii$) removes most if not all spurious (non-astrophysical) events and many signatures that are too noisy to characterize at the level required for validation.  Criterion ($iv$) is used to verify the periodicity of the transits.  We added criterion ($v$) because, as shown in Figure \ref{fig:perdist}, the ratio of the abundance of identified FPs to candidate transiting planets in multis is twenty times as large for orbital periods $P < 1.6$ days as it is for  $P > 1.6$ days. This implies that the EB and planet priors input into a statistical analysis to estimate the relative probability of transiting planet and EB sources are skewed towards EBs at short period and towards transiting planets at long periods.  (See Appendix B for further discussion of the period distributions of various classes of transit-like signatures.) While {\it bona fide} planetary transits with high impact parameter can be V-shaped, a far larger fraction of stellar eclipses have this form, inspiring criterion ($vi$).

Note that the six criteria listed above exclude some well-known \ik planets, including the multiples Kepler-10 \citep{Batalha:2011, Fressin:2011} and Kepler-42 \citep{Muirhead:2012a}, each of which has only one planet candidate with $P > 1.6$ days, and Kepler-47 \citep{Orosz:2012}, which is a circumbinary system. (Transits of circumbinary planets, i.e., planets on P-type orbits, are irregularly spaced and of varying shapes, and have not been correctly identified by the \ik automated transit search. We thus exclude 
circumbinary planets, which can be considered a separate class, with 
 poorly-measured multiplicity because they are not 
discovered in a standardized manner.) However, these omissions are acceptable in the context of this work because the vast majority of \ikt's candidate multis meet all six of the above criteria, and our goal is to examine systems that are relatively easy to validate with a high level of confidence rather than to address all viable candidate multiple planet systems.

\section{Numerical Estimates of FPs for \ik Candidates}

In order to validate a large sample of \ik candidates as planets with a high level of confidence, we need to show that the expected number of FPs in said sample is much smaller than the size of the sample.  The statistical analysis presented above treats all \ik targets as equally good for searching for planet candidates, and searching for individual candidates is assumed to be independent of the presence of other candidates.  Unfortunately, the dataset and search processes are not so simple.  We are thus forced to make certain approximations that cannot be rigorously justified.  In doing so, we note that because it is more important to bound the false detections rate from above rather than to estimate its value precisely, we favor assumptions that overestimate the predicted FP rate in comparison to the number of multis included in the sample.  According to Equations (\ref{eq:2fp}) -- (\ref{eq:2p2fp}), which we use to compute statistical estimates of the numbers of FPs, this means we should avoid overestimating $n_t$.
  
Paper III describes the ensemble of planetary candidates.  Including all viable candidates as well as identified FPs and false alarms (FAs, which are KOIs with S/N $<$ 7.1 indicating that may well be spurious, i.e., not astrophysical in origin, or were produced by intrinsic stellar variability or show fewer than three putative transits; see Paper III for details) yields a grand total of 466 multis, the characteristics of which are presented in Table 3 
of Paper III. These active candidates, FPs and FAs were found by searching through the light curves of a total of  190,751 \ik targets observed at some time during Q1 -- Q8.  After applying the six criteria listed in Section 3, we have: $n_1 = 1303$, $n_2 = 272$, $n_3 = 96 $, $n_4 = 40 $, $n_5 = 10 $, $n_6 = 2 $ (so $n_m = 420$, $n_k = 1723$ and $n_c = 2357$).

Because of variations in photometric noise of a target, stellar size, magnitude and amount of time observed, the targets provide a very heterogeneous sample with regards to planet detectability. Thus, even if transiting planets were randomly distributed among the \ik  targets, we would expect some degree of clustering (a bit less uniformity, i.e.,  perhaps a few more multis, than a random distribution) in detected planets. This implies that the appropriate value of $n_t$ for our statistical estimates of the numbers of FPs in multis is smaller than 190,751. Unfortunately, it is not possible to make any sharp cuts in stellar host properties among the planet candidates, but we can examine trends in order to produce a more appropriate estimate of $n_t$.

Table \ref{tab:targetcount} lists number of targets and the number of multis as functions of the number of quarters during the Q1 -- Q8 interval for which the target was observed.  The discovery rate for multis (broadly defined to include FPs and FAs), given in the last column of Table \ref{tab:targetcount}, decreases substantially when fewer than five quarters of observations exist.

Giant stars ($\log{g} < 3.5$) occupy $\sim 10\%$ of the \ik target list but host only  $\sim 1\%$ of the planet candidates. Planets transiting giant stars  produce smaller transit depths, hence, giant stars are a priori poor targets for transiting planet searches. The fraction of stars  listed as unclassified in the \ik input catalog (KIC) that host planet candidates is less than half that of stars classified as dwarfs, suggesting that a majority of the KIC unclassified stars are also giants.

\begin{figure}
\begin{center}
\includegraphics[scale=0.5]{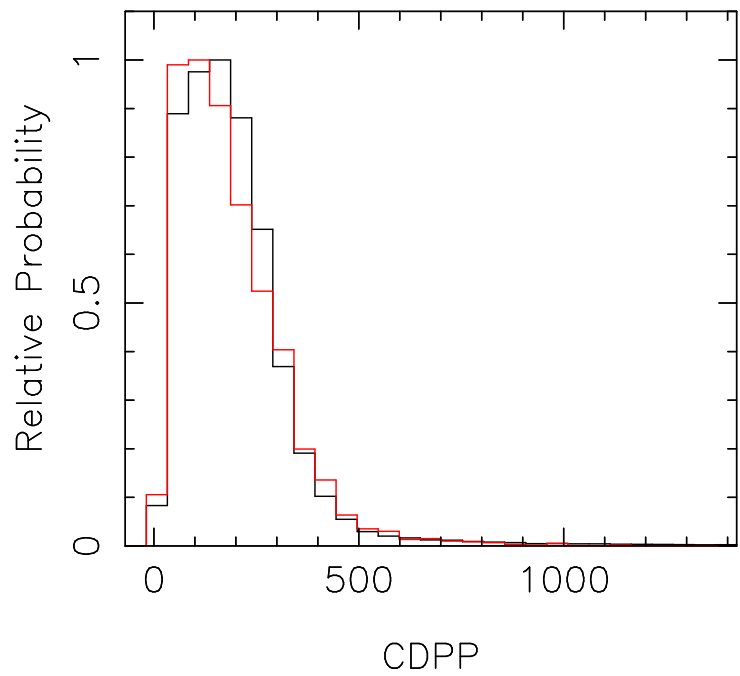}
\end{center}
\caption[]{Comparison of 3 hour CDPP for all \ik targets (black) versus those targets showing planet candidates or FPs (red). As expected, transit-like signals are preferentially found around targets with less noise. However, the differences between the two distributions are small.}
\label{fig:cdppdist}
\end{figure}

To measure photometric suitability of a target for detection of planet candidates, we use the combined differential photometric precision (CDPP), which is a measure photometric noise on timescales related to the transit duration.   A CDPP of 20 ppm for 3-hour transit duration indicates that a 3-hour transit of depth 20 ppm would be expected to have an S/N = 1 \citep{Christiansen:2012}.  We use the 3-hour estimate from a Q1 -- Q10 \ik pipeline run. (Three hours  is a close match to the typical transit duration of our sample.)  Figure \ref{fig:cdppdist} shows a histogram of the CDPP distribution.  The black line shows the entire \ik target Q1 -- Q8 sample and the red line shows the distribution for those targets from the Q1 -- Q8 planet search with currently viable planet candidates (singles or multis).  We calculate a median CDPP value of 160 ppm for the planet candidate hosts, compared with 175 ppm for all \ik targets.

Based on our results shown in Table \ref{tab:targetcount} and Figure \ref{fig:cdppdist}, as well as the paucity of planet candidates orbiting giant stars, we choose to estimate the effective size of our target pool, $n_t$, as follows: The median CDPP of targets hosting planet candidates is 160 ppm. We counted 70,008 targets classified as dwarfs ($\log{g} > 3.5$) in the KIC, observed in the majority of the eight quarters searched, and with CDPP $<$ 160 ppm.  We then doubled this number so that the median photometric variability within the effective target pool is equal to the median photometric variability of \ik planet candidate hosts, yielding $n_t = 140,016$. Thus, the value of $n_t$ used for our FP estimates is equal to twice the number of targets that were observed for at least five of the eight quarters searched for transits and that have $\log{g} > 3.5$. Note that by excluding unclassified stars, some of which are dwarfs,  we slightly underestimate the effective number of targets, which is conservative in the sense of overestimating the numbers of FPs in multis because $n_t$ appears in the denominators of Equations (\ref{eq:2fp}) -- (\ref{eq:2p2fp}).

\begin{deluxetable}{cccc}
\tabletypesize{\scriptsize}
\tablecaption{Targets and Multis Discovery Rate}
\tablewidth{0pt}
\tablehead{
\colhead{\# of Quarters}   & \colhead{\# of Targets} & \colhead{\# of Multi-KOIs} & \colhead{\%}
}
\startdata
1 &  5648  &   0 & 0.00 \\
2 &  4738  &   1 & 0.02 \\
3 & 5366  &   1 & 0.02 \\
4 & 9869  &  11 & 0.11 \\
5 & 11404  &  23 & 0.20 \\
6 & 11027 &   22 & 0.20 \\
7 & 55851 &   156 & 0.28 \\
8 & 86848 &    252 & 0.29 \\
 \\
 Total & 190751 & 466 & 0.24 \\
\enddata
\label{tab:targetcount}
\end{deluxetable}

The observed candidate abundances used as input for our estimates of the number of as yet unidentified FPs in \ik multis in Section 4.1 are listed above and given in the top row of Table \ref{tab:multicount}. For this analysis, we conservatively assume a true planet fraction of ${\cal P}_1 = 0.9$ for the singles, a value that corresponds to a significantly larger fraction of (non-planetary) FPs than estimated by \cite{Fressin:2013} and \cite{Santerne:2013}; cf.~Appendix C. The bottom two rows of Table \ref{tab:multicount} give input abundances for our test analyses of ``dirty'' samples presented in Section 4.2 that estimate the abundance of identified FPs in multis using the identified FP rate among targets that show a single transit-like pattern.

\begin{deluxetable}{cccccccccl}
\tabletypesize{\scriptsize}
\tablecaption{Multiplanet Counts for Statistical Studies}
\tablewidth{0pt}
\tablehead{
\colhead{$n_t$} & \colhead{$n_1$}  & \colhead{$n_2$} & \colhead{$n_3$} & \colhead{$n_4$} & \colhead{$n_5$} & \colhead{$n_6$} & \colhead{objects included}
}\startdata
 140\,016 & 1303 & 272 &  96 &  40 &  10  &  2 &  statistics for validation \\
 - & 1520 & 284 & 102 & 43 & 13 & 3 & all candidates\\
 140\,016 & 2499 & 300 & 104 &  43  & 13  &  3 & all candidates + FPs \\
140\,016 & 2637 & 300 & 104 &  43  & 13  &  3 & all candidates + FPs + EBs\\
\enddata
\label{tab:multicount}
\end{deluxetable}

Our goal is to compute conservative estimates of the expected number of false positives in \ik multis, but Equations (\ref{eq:2fp}) -- (\ref{eq:1p2fp}) and (\ref{eq:2p2fp}) 
include the quantity $n_{\rm fm}$, the total number of (not yet identified) false positives lurking in \ik multis. 
 We resolve this need for assuming a result prior to obtaining it by making conservative assumptions on the value of $n_{\rm fm}$ and then checking that these assumed values are higher than the expected values.  This can be done by iteration, which for unidentified FPs in multis quickly converges to $n_{\rm fm} <$ 3 for  ${\cal P}_1 \geq$ 0.9. 

\subsection{Sample Considered for Validation}
Inserting the numerical values given above into Equations (\ref{eq:2fp}) -- (\ref{eq:2p2fp}), we get the expected numbers of the various systems of multis containing unidentified FPs given in  Table \ref{tab:fpest}. These values add up to a total of 1.57 FP planet candidates (counting systems with 2 FPs twice) expected in 2-candidate systems and 0.52 FP in systems of 3 or more candidates; our estimates decrease slightly faster than $1 - {\cal P}_1$ for values of ${\cal P}_1 > 0.9$.  
 These final numbers are to be 
compared to the observation that there are 272 candidate 2-planet systems, with a total of 544 planet 
candidates, and 148 candidate systems of 3 or more transiting planets, with a total of 510 planet 
candidates satisfying the six criteria listed in Section 3. Thus, \emph{for the assumptions that we have made, this sample of 1054 planet  candidates in multis should be $\sim$ 99.8\% real planets.}  But to compensate for the issues raised in Paper I and in Sections 5 and 6, we don't simply validate all of these candidate multi-planet systems, but rather we subject them to additional scrutiny.

\begin{deluxetable}{ccccc}
\tablecolumns{5} \tablewidth{0pc} 
\tablecaption{Statistical Estimates of Unidentified False Positives in Multis}
\tablehead{ 
	\colhead{class (formula)} &
	\colhead{Expected Number (for ${\cal P}_1 = 0.9$)} } 
\startdata
2 FPs (Eq.~\ref{eq:2fp}) & 0.063  \\
3 FPs (Eq.~\ref{eq:3fp}) & $2.0 \times 10^{-5}$  \\
1 planet + 1 FP (Eq.~\ref{eq:1p1fp}) & 1.447  \\
1 planet + 2 FPs (Eq.~\ref{eq:1p2fp}) & $5.3 \times 10^{-4}$ \\
$\ge 2$ planets + 1 FP (Eq.~\ref{eq:2p1fp}) & 0.517  \\
$\ge 2$ planets + 2 FPs (Eq.~\ref{eq:2p2fp}) & $1.9 \times 10^{-4}$  \\
\hline\
Total FPs (\# of false candidates) & 2.09 \\
\enddata
\label{tab:fpest}
\end{deluxetable}

\subsection{Statistics of False Positives Identified in Singles and Multis}

The fraction of  \ik planet candidates that are false positives is likely to be quite low, but the FP rate is much higher among the entire sample of transit-like signatures that have been identified within the \ik data.  The share of known FPs among the entire ensemble of KOIs identified using the first two years of \ik data (not just the vetted KOIs that have been promoted to planet candidates) exceeds 27\%, and the \ik eclipsing binary catalog (Kirk et al., in preparation) is more than half as long as the entire KOI list used for our study.  Comparison of the numbers of FPs identified in singles and multis can, in principle, provide a test of the statistical argument for the relative FP rates of singles and multis in the remaining ensemble of planet candidates that forms the basis of our planet validation technique.

Unfortunately, while some portions of the search for transiting planet signatures within \ik data are automated and can be well-characterized, promotion of a threshold crossing event to a KOI involves subjective decisions (Paper III). 
Thus, the current KOI list is biased in ways that make it unsuitable for highly accurate testing of our statistical analysis. These biases are in both directions, so we can't say which sign the errors will be. Nonetheless, we proceed to make two tests of the  predictions from Equations (\ref{eq:2fp}) -- (\ref{eq:2p2fp}), one of which includes  low depth EBs as FPs, the other does not, to see whether or not these formulae yield roughly correct results.

The dataset that we use to test our statistical framework for estimating FP rates in multis is the \emph{unvetted} KOI list (i.e., planet candidates + identified FPs) through Q8 with very low amplitude events removed.  Specifically, we apply criterion ($i$) of Section 3 as well as a weak version of criterion ($iii$): signal-to-noise ratio S/N $>$ 7.1 (through Q14).  Events with  S/N $<$ 7.1 through Q14, i.e., false alarms,  comprise $\sim$ 2\% of both the singles and the multis (Section 5.2 of Paper III); these abundances are not relevant to our statistical study because searching for additional planet candidates has at times been more aggressive for targets with a candidate already identified, leading to extra spurious events being added to the planet candidate list\footnote{A few spurious events may remain as active planet candidates with S/N slightly above 7.1 and be treated as such for the statistical tests presented in this subsection. However, as noted in Section 5, the stricter requirement of S/N $> 10$ implies that it is unlikely that any of these events qualify for our primary statistical study whose results are presented in Table \ref{tab:fpest}. And our validation protocol makes it even less likely that we validate any spurious signals as exoplanets.}. This yields a  list that contains $n_k^* = 2962$ targets, including $n_m^* = 463$ with multiple sets of transit signatures (third row of  Table \ref{tab:multicount}). None of the candidates in the 59 systems with 4 or more planet candidates have been identified as FPs,  two of the 3-KOI systems have a single FP, 
six of the 2-KOI systems have both KOIs identified as FPs, 
 and 12 have one of the pair identified as an FP
, totaling 26 FPs out of a total of 1167 KOIs (2.2\%).  In contrast, 1017 of $n_1^* = 2499$ single KOIs have been identified as FPs (41\%), implying an FP rate \emph{for the unvetted sample of KOIs} among the singles that is more than ten times as large as that among the multis and a value of ${\cal P}_1^* = 1480/2499 = 0.5922$. An additional 138 targets have been classified as detached (non-contact) eclipsing binaries of small ($< 2\%$) depth; if these are counted as false positives (bottom row of  Table \ref{tab:multicount}), then the FP rate for singles increases to 44\% (${\cal P}_1^{**}$= 1480/2637 = 0.5612).

We present a comparison between observations and predictions in Table \ref{tab:stattest}. We choose $n_t^{**} = n_t^* = n_t =$ 140,016 and estimate the expected number of FPs in multis iteratively as $n_{\rm fm}^*$= 24.9 and $n_{\rm fm}^{**}$= 29.4. Numerical estimates for specific classes of FPs among unvetted systems of multis containing FPs that are calculated using Equations (\ref{eq:2fp}) -- (\ref{eq:2p2fp}) are presented in successive rows of  Table \ref{tab:stattest}.  
These tests show excellent agreement between predicted and observed numbers of FPs in multis.

\begin{deluxetable}{ccccc}
\tablecolumns{5} \tablewidth{0pc} 
\tablecaption{Tests of Statistical Estimates using Identified False Positives}
\tablehead{ 
	\colhead{class (formula)} &
	\colhead{Observations} &
	\colhead{Predictions w/o EBs} &
	\colhead{Predictions with EBs} } 
\startdata
2 FPs (Eq.~\ref{eq:2fp}) & $6$ & 3.89 & 5.03 \\
3 FPs (Eq.~\ref{eq:3fp}) & 0 & 0.0097 & 0.014 \\
1 planet + 1 FP (Eq.~\ref{eq:1p1fp}) & 12& 12.98  & 14.66 \\
1 planet + 2 FPs (Eq.~\ref{eq:1p2fp}) & 0 & 0.042 & 0.054 \\
$\ge 2$ planets + 1 FP (Eq.~\ref{eq:2p1fp}) & 2 & 3.99  & 4.50 \\
$\ge 2$ planets + 2 FPs (Eq.~\ref{eq:2p2fp}) & 0 &  0.013 & 0.017 \\
\hline\
Total identified FP candidates in multis & 26 &  24.9 &  29.4 \\
\enddata
\label{tab:stattest}
\end{deluxetable}

\section{Planets, Host Stars, and Planetary Systems}
 
Various caveats to the applicability of the statistical analysis presented above are discussed in Section 2.2 of Paper I (``three'' in the first sentence of that section should read ``two").  Background stars are more common near the galactic plane, leading to a higher probability of finding background eclipsing binaries (BGEBs), so chance-alignment blends are not completely random. The overall effects of this skewed distribution  are likely to be small because they average out in the same manner for single FPs in multis as for FPs among targets with only one identified pattern of transits (singles), but suggest a bit of extra caution when assessing candidate multi-planet systems at low galactic latitude. Given the similarity in the latitude distributions of \ik targets and planet candidates evident in Figure 12, our incorporation of the concentration of background stars towards the galactic plane into criterion ($vii$) of our validation protocol presented in Section 7 is sufficient to address this concern.  We consider the most likely interlopers in our sample below.

A signal in a \ik light curve that is interpreted as a planet candidate could be produced by a variety of processes, which we categorize as follows:

(I) a planet transiting the \ik target star

(II) a planet transiting a stellar companion to the \ik target star

(III) a planet transiting a star not physically bound to the \ik target star

(IV) an eclipsing binary star system or other astrophysical phenomenon

(V) false alarms resulting from noise, processing errors, etc.

A multi with all candidates in category (I) is a clean planet detection, whereas any candidates in categories (IV) and (V) are clearly non-planetary FPs/FAs.  But what of signals in categories (II) and (III), and of multis composed of mixtures of planets in categories (I) -- (III)?  Two key derived properties of an individual planet, size and the radiation flux from its star that it intercepts (insolation, and hence equilibrium temperature), require knowledge of the star being orbited, whereas the orbital period of the planet (Doppler shifted to the rest frame of our Solar System, as discussed in \citealt{Lissauer:2011a}) does not.  The derived age of the planet and distance of the planet from the Solar System depend on the stellar system in which the planet orbits, hence mistaking (II) for (I) does not affect these quantities\footnote{The presence of a bound companion can affect the derived age of a system and its  distance from the Solar System, leading to errors in estimated properties regardless of whether the planets orbit the primary or the secondary.}.  In contrast, from a planetary dynamics standpoint, the most important issue is whether or not all members of a multi orbit the same star (see \citealt{Fabrycky:2013} and Section 6), although planetary sizes are also of interest since masses are correlated with sizes.  And for \ikt's statistical census, it is critical to identify the star or at least the stellar system to which the planets are bound.

Planet validation must demonstrate that the statistically likelihood of a signal being an FP is  quite small compared to that of it being produced by a planetary transit.  Thus, the probability of each candidate being in category (IV) or (V) must be very small.  Stochastic noise is expected to produce no more than one spurious transiting planet signature with S/N $>$ 7.1 over \ikt's entire four year mission \citep{Jenkins:2002}. By employing the more stringent  S/N $>$ 10 threshold (criterion $iii$), we expect that few if any of these false alarms are included in our statistical analysis.  Our validation protocol also requires inspection of all individual folded light curves in the vicinity of transits (Paper III), so it is unlikely that any such FAs contaminate our sample of validated planets. The worst misidentification would be validating a target for which no signals are of planetary origin as a planetary system. The primary expected source of such entirely FP multis are pairs of eclipsing binaries, with an expectation value for the entire sample being considered that is given by Equation (\ref{eq:2fp}) to be significantly smaller than unity (Table \ref{tab:fpest}). Validating members of multis that do not correspond to planets is almost as bad. Here the estimated numbers are quantified in Equations (\ref{eq:1p1fp}) and (\ref{eq:2p1fp}); one or two such FPs may well occur in our listing of validated planets, or perhaps there are three or none, but in any case the abundance of FPs among our sample is expected to be well below 1\%, making our listing of validated planets in multis of very high reliability by the standards of tabulations of ``known'' exoplanets. 

 According to the calculations of \cite{Fressin:2013} as modified by \cite{Santerne:2013} to account for secondary-only eclipsing binaries (occultations), $88.7\pm1.1$\% of the \ik planet candidates listed in \cite{Batalha:2013} are expected to be planets transiting the target star (category I above).  The next largest grouping, representing $\sim 6.6\%$ of the candidates, are planets orbiting bound companions to \ik targets (category II above). Eclipsing binaries (category IV above) are expected to account for $\sim 4.2\%$ of the candidates, with $\sim 2.2\%$ representing physically-associated EBs  and $\sim 2.0\%$  chance-alignment blends with eclipsing binaries. Just $\sim 0.6\%$  are chance-alignment blends with transiting planets (category III above). Thus, planets orbiting physically-associated stars are expected to be an order of magnitude more common among planet candidates than planets orbiting background stars.  In contrast, FPs from background eclipsing binaries are expected to be almost as common as those from EBs that are physically associated with the target (either in a triple or higher multiplicity star system, or a secondary-only occultation by the target star), and  for planets smaller than Neptune, which represent most of the planets that we validate in our study, background EBs dominate bound EBs by more than an order of magnitude \citep{Fressin:2013, Santerne:2013}.  
 
 Although the  probability of a transit-like signal being produced by a background eclipsing binary is small when averaged over all targets, several factors lead to large variations from target to target.  The FP estimates computed by \cite{Fressin:2013} and \cite{Santerne:2013} represent averages over a very heterogeneous sample and do not account for planetary multiplicity.  The {\it a priori} likelihood of a chance-alignment blend for a given target depends on the population of background stars, which is more than an order of magnitude larger at the lowest galactic latitudes observed by \ik than at the highest latitudes \citep{Morton:2011}. This likelihood also depends on the magnitude of the target star, depth of the transit signal, and the area around the target to which the transit can be localized, which varies by more than two orders of magnitude.  The above mentioned factors are all positively correlated (to varying degrees) among planet candidates of a given target.  Such correlations do not affect our estimates of the number of multis that include a single FP, but they imply that the expected number of multiple FPs is higher than predicted by Equations (\ref{eq:2fp}), (\ref{eq:3fp}), (\ref{eq:1p2fp}) and (\ref{eq:2p2fp})\footnote{Indeed, the tests presented in Section 4.2 show a small excess of identified double FPs and a small deficit of single FPs in multis relative to our predictions (Table \ref{tab:stattest}), although the differences from predicted values are not statistically significant.}. Fortunately, these equations give estimates substantially lower than the corresponding estimates for single FPs in multis given by Equations (\ref{eq:1p1fp}) and (\ref{eq:2p1fp}), so our general conclusion that the FP rate is very small in multis still holds.  \emph{Nonetheless,  the huge variations in localization of transit position imply that centroids computed using pixel-level data obtained by \ik must be examined for us to locate the source of the transit signatures on the plane of the sky sufficiently well for the validation of planets via multiplicity to be robust.}

 Some of the \ik candidate multi-planet systems could well be planetary systems orbiting stars other than the nominal target star (either chance-alignment blends or fainter physical companion stars).  The number of such planetary systems orbiting chance-alignment blends is likely to be small because, as stated above, \cite{Fressin:2013} estimate an order of magnitude fewer blends with background planets than with planets transiting companion stars. Moreover, this differential is likely to be even larger for multis  because magnitude differences are typically larger between chance-alignment blends than between physically-associated stars, which are essentially the same distance from the observer, and most multi-planet systems contain small planets that would be difficult to detect around faint stars \citep{Latham:2011}. 
 \cite{Fressin:2013} estimate that chance-alignment blends from background transiting planets produce a bit less than $1\%$ of the signals classified as \ik planet candidates. Thus, we expect that few if any of the multis that we validate as planetary systems are planetary systems that are not orbiting either the \ik target star or a star that is gravitationally bound to the \ik target.  Moreover, we do not consider them to be false positives because they are true planets and true planetary systems.  Note that as we are not requiring spectral analysis of planet-host stars for  validation, substantial errors in estimates of planet sizes may be present in our validated planet lists even in cases where the planets orbit the \ik target - {\it caveat emptor}!

\section{Multiple Stars and Split Systems}

The most insidious interlopers in our sample of validated planets are likely to involve stars that are physically bound to the \ik target.  These include planetary systems that orbit a secondary to the target, implying that the planets are significantly larger and intercept significantly less stellar flux than estimated assuming that they orbit the primary, as well as ``split systems'' containing two stars each with transiting planets on S-type orbits or one star with transiting planet(s) and two (typically fainter) companion stars that form an eclipsing binary.  Systems of planets all of which orbit a physical companion to the \ik target are more difficult to detect than those around the target because their transit signals suffer a dilution exceeding 50\%, but nonetheless are likely to be more common than those orbiting chance-alignment, unbound, background stars (blends). 

\subsection{Numerical Estimates for Uncorrelated Orientations}

The abundance of split systems containing one or more planets orbiting the \ik target together with a binary companion star that has transiting planets of its own or is eclipsed by a third star in the system can be estimated if additional assumptions along the line of those in Section 2.4 of Paper I are made. Specifically, we assume that:

\noindent (1) Half of the \ik target stars are binaries, which is approximately correct. According to \cite{Raghavan:2010} the fraction of star systems that are multiples is 44\%, but because of triples and systems of higher multiplicity, the mean number of stellar companions to a given primary star exceeds 0.5. Also, the \ik target list is a magnitude-limited sample, so it is somewhat biased in favor of binaries that consist of stars of comparable brightness relative to the volume-limited sample of \cite{Raghavan:2010}. A counteracting bias is that if the two stars are sufficiently distant from one another to have been viewed as separate objects during target selection, the ``crowding'' may have been cause to reject both from the \ik target list.  The \ik target list includes a wide range of stellar spectral types, but is primarily composed of sunlike stars, which were the subject of the study  of \cite{Raghavan:2010}. 
 
\noindent (2) Planets are as easy to detect (at S/N  $>$ 10) in the \ik data around stars in binaries as around single stars. This yields an overestimate of the number of detectable planets orbiting stars in binaries, both because of dilution and the fact that many of the secondary stars would be too faint and {\it a priori} poor for planet detection to have been included in the \ik target list if they did not have a bright companion.

\noindent (3) There is no correlation of the orientations of the two orbital planes of planetary systems orbiting bound stars.  This assumption  is expected to underestimate the number of split systems, because there is likely to be some tendency of alignment of orbital planes.

\noindent (4)  Stars in binaries are equally likely to host planets  as are single stars.  We expect this assumption to lead to an overestimate of the fraction of observed transiting planets in binaries, because binary companions destabilize some planetary orbits and, as discussed in Section 8, the sample of \ik multis does not show evidence for widespread stellar binarity.

As roughly 50\% of solar-mass stars have a stellar companion,  there are $\sim 3n_t/2$ stars being searched for planets 
and $\sim n_t/2$ pairs of stars that could possess split systems.  The total number of stars harboring one or more planet candidates is $n_k + n_B$, where the term $n_B$ accounts for the targets with two stars each possessing at least one planet candidate. It follows that the fraction of all stars surveyed (including binary companions) that harbor at least one planet candidate is $(n_k + n_B) / (3n_t/2)$.   The estimated fraction of binary stars for which each of the  stellar components harbor a close companion whose eclipse, transits or occultations produce a  planet candidate is the square of this fraction, assuming uncorrelated occurrence of planets around the two stars.  Thus the number of binary stars, $n_B$, having a planet candidate around each stellar component is given by:
  
\begin{equation}
\label{eq:Bp}
{\rm binary~stars,~each~with~planet~candidate(s):}~n_B = \biggl(\frac{n_k + n_B}{3n_t/2}\biggr)^2\frac{n_t}{2}  = \frac{2(n_k+n_B)^2}{9n_t}.
\end{equation}
Solving Equation (\ref{eq:Bp}) iteratively yields $n_B \approx 4.74$ mixed  systems. For ${\cal P}_1 = 0.9$, 80\% of such split systems with two planet candidates and more than 90\% with three or more planet candidates are composed entirely of planets.

We employ an alternative methodology to estimate the number of such split systems containing an EB FP, again neglecting any possible correlations in orbital planes. We first estimate the number of false positives caused by EBs for targets in triple (or higher multiplicity) star systems, $n_{\rm fT}$, by summing the contributions identified in \cite{Fressin:2013} and \cite{Santerne:2013} and multiplying by 2357/2222 to account for the larger size of the sample of planet candidates that we are analyzing, yielding the estimate $n_{\rm fT}$ = 27.  The expected number of such contaminants is given by:

\begin{equation}
{\rm transiting~planet(s)~+~bound~EB~FP:}~\frac{n_{\rm fT} n_k}{n_t}   .
\end{equation}

Numerical estimates based on Equations (\ref{eq:Bp}) and (10) suggest that $\sim$ 4 or 5 mixed planetary systems and $\sim 0.33$ combinations of a bound EB and a planetary system (the latter being included in Eqs.~\ref{eq:1p1fp} and \ref{eq:2p1fp}) are present in the sample of \ik multis. Thus, under the four assumptions above, mixed planetary systems are estimated to represent $\sim 1\%$ of the sample and bound EBs occur in $\ll 1\%$.  But note that if orbital planes are correlated, or if planet occurrence is correlated, the number of split systems could be much greater than estimated from these statistical considerations alone.

According to \cite{Fressin:2013} and \cite{Santerne:2013}, who neglected planetary multiplicity, transit signals from planets orbiting bound stellar companions to \ik targets are expected to be $\sim 3$ times more common among planet candidates than transit signals from EBs that are physically bound to the target (implying a triple or higher multiplicity star system), and for candidate planets smaller than 6 $R_\oplus$, which represent an overwhelming majority of the planets that we validate in our study (Section 7, especially criterion $x$), the likelihood of planets orbiting a companion star exceeds bound EBs by more than an order of magnitude.  These results of \cite{Fressin:2013} reinforce our general conclusions that the number of such stellar triple interlopers (with one planet candidate being an eclipsing binary star system and therefore by all accounts an FP) is likely to be quite small, but the number of split systems composed of true planets with different (albeit bound) stellar hosts is expected to be larger.  


In sum, the percentage of EBs in multis is expected to be far smaller than that in singles.  The percentage of transit signals in multis that result from planets orbiting stars other than the nominal \ik target star is expected to be significantly less than in singles because small planets, which make up a higher percentage of the multis than of the singles, are more difficult to detect than are large planets when the light from their star contributes a minority of the photometric signal observed by \ikt. Nonetheless, these planetary interlopers are likely to be far more common than EBs in multis.  

\subsection{Stability and Split-Multis}

\cite{Fabrycky:2013} estimated the number of 
``split-multis'', i.e., multis that do not represent  
planets orbiting the same star, which include chance alignment blends as well as the physically-bound split systems discussed above. Their study employed a statistical analysis of the number of candidate multis with orbital periods that, together with masses estimated from observed sizes, imply instability for planets orbiting the same star. They noted that two multis in the \cite{Batalha:2013} cumulative catalog, the 
3-candidate system KOI-284 and the 4-candidate system KOI-2248, would clearly be unstable for any reasonable 
planetary masses, but that all other multis are stable for sufficiently small orbital eccentricities and the assumed empirical mass-radius relationship that we reproduce in Equations (11) and (12) in Appendix D, and estimated the number of split-multis that would 
produce exactly two cases that appear unstable because of small period 
ratios. \cite{Fabrycky:2013} found that split planet pairs in multis (each combination of two planet candidates of the same \ik target is considered a single planet pair, so each $i$ candidate system has $i(i-1)/2$ planet pairs) represent 1.1 -- 13\% of the sample of planet pairs in multis presented by \cite{Batalha:2013}, with the quoted range corresponding to  the 95\% credible level, analogous to the 2$\sigma$ range. These estimates are well above our estimates for FPs in multis, so we consider the population of unstable multis further in this section.  Note that 
split-multis can be composed exclusively of planets, as is  the 
case for KOI-284 (Section 9.1; see also Paper I and Bryson et al., in preparation).

The candidate pool that we consider for our statistical analysis, i.e., those satisfying the criteria listed in Section 3, contains no unstable planet candidate pairs apart from the two noted by \cite{Fabrycky:2013}, despite the larger size of this new sample (see Appendix D). The KOI-2248 system fails our validation criteria because only one of the candidates, KOI-2248.01, both passes
our  centroid tests for localization of the transit signal on the plane of the sky and has $P > 1.6$ days. 
The planet candidate with a period similar to that of 2248.01,  2248.04, has too much centroid 
scatter to clearly distinguish whether its transit signature is from the nominal \ik target or a neighboring star that is about 4$\arcsec$ from the target star and 7\% as bright as the \ik target in the \ik bandpass. 
Therefore 2248.04 did not even pass the weaker form of transit localization that we would accept for validation of additional planet candidates of targets with two or more candidates that we are validating as planets.

In contrast to KOI-2248, KOI-284 passes all of our standard validation tests.  But with only one such system, the population of split-multis among our validated planets  does not lend itself to robust results from the type of analysis performed by \cite{Fabrycky:2013}. Moreover, since all three candidates in KOI-284 appear to be planets, we do not consider the split-multis results of \cite{Fabrycky:2013} to  cast doubt upon our validation procedure.

\section{Criteria for Planet Validation}

Our goal is to validate a large number of \ik multis as planets and to have few contaminants slip into our validated planet list.  In order to achieve these goals, we require that to be validated as  planets, at least two of a  target star's planet candidates meet all six requirements for the statistical analysis stated in Section 3 as well as the following additional criteria:

($vii$) Centroid analysis  \citep{Bryson:2013} indicates that  the location of the transit signal on the plane of the sky is coincident with that of the \ik target star within the narrow tolerance (typically $\lesssim 1\%$ of the area of the \ik aperture) specified in Section 5.6 of 
 Paper III and that the probability of the transit signal originating on another star is $< 1\%$. 

($viii$)  For orbital periods 1.6 days $< P < 4$ days, the signal-to-noise ratio requirement is increased to S/N $>$ 15 (through Q14). This requirement was added to insure a robust measurement of the transit shape because in this period range the ratio of the abundance of EBs to that of transiting planets is larger than it is for longer periods. 

($ix$) Periods of neighboring planets are not so close as to suggest that the configuration being validated is dynamically unstable (see Appendix D).  

($x$) The largest estimate of planet size (2$\sigma$ upper bound) is $R_p < 9~R_\oplus$. This criterion is relaxed to $R_p < 15~R_\oplus$ if  another planet candidate is orbiting sufficiently close that it either would not be in a stable orbit or would have larger transit timing variations than observed if the mass of the $9~R_\oplus < R_p < 15~R_\oplus$ object exceeded 13 times the mass of Jupiter (see Appendix D). These limits were chosen to exclude stars and brown dwarfs, the minimum size for which is $9~R_\oplus$ \citep{Chabrier:2000}.




As stated above, we only validate systems in which two or more planet candidates meet all of the above requirements.  Fewer than 1 in 300 \ik targets meet these conditions and are thereby validated as multi-planet hosts by our methods.  The probability that an additional candidate of one of these targets is actually a planet bound to the same host star is quite high, so we relax one of our validation requirements for such planet candidates.  Specifically, we impose a weaker form of criterion ($vii$) that allows for a somewhat larger uncertainty in the position of the transit signal on the plane of the sky (but that nonetheless localizes the transit source to an area $< 10\%$ that of the \ik aperture that includes the target star; see 
 Paper III).  We also validate one candidate that fails one of our standard validation criteria but for which there is strong additional evidence for it being a planet orbiting the same star as six planets validated by our standard techniques; see Sections 11 for details.

Our statistical analysis predicts extremely few FPs in multis, but astrophysical uncertainties in the planarity of multiple star/planet systems imply that we cannot make a good estimate of the numbers of FPs that may have passed our tests and contaminate our list of validated planets. Specifically, FP rates from chance-alignment blends of a background signal in combination with a signal from the \ik target 
are a quantifiable unknown. In contrast, the expected number of FPs caused by a combination of one or more planets transiting the \ik target and a physical companion pair of eclipsing binary stars can only be quantified if one knows or makes assumptions about correlations between the orbital plane of a planetary system around one star with the plane of a close binary that is physically bound to the planet-hosting star. If one assumes no such correlations, FPs from physical triples can be estimated as a quantifiable unknown (which has a very low expectation value; see Section 5); assuming strict coplanarity also gives quantifiable numbers, but an intermediate situation is likely to be correct, leading to an unquantified (and with our present astrophysical knowledge unquantifiable) unknown.  Similar divides are present among planets that are not all orbiting the target star (Section 5).  We expect the largest contaminants to our ensemble of validated planets presented in this work and in Paper III to be unquantified unknowns, with very small numbers of quantified unknowns (and also of truly unknown unknowns, but this of course cannot be proven).  Despite these caveats,  we are confident that more than 99\% of the objects that we validate as planets should indeed be true planets, and that in well over 90\% of the cases, all of the validated planets that we associate with a given \ik target  orbit the star that we associate with them, i.e., are true planetary systems about this specified \ik target.   

\section{Search for Stellar Companions}

\subsection{Spectroscopic Analysis}

For our spectroscopic search for secondary stars, we obtained and examined high resolution spectra of 259 of the 466 host stars of
multis. These data were taken with HIRES spectrometer on the Keck I telescope using the observing setup of the CPS group \citep{Marcy:2008}, without the iodine cell in the light path. They provide  a wavelength coverage of 360 -- 800 nm  with a resolution of R = 55,000 and have a S/N per pixel of 40 (or better) at 550 nm, corresponding to a S/N = 85 per resolution element. All spectra were taken 
with the C2 decker, which projects to $0\arcsec.87 \times 14\arcsec.0$ on the sky, offering $5\arcsec$ of sky spectrum on either side of 
the stellar spectrum, permitting subtraction of moonlight, if any.   We 
proceeded to analyze the spectra for evidence of a secondary spectrum, 
as described in detail in \cite{Marcy:2014}.

For these 259 host stars of multis, we searched for spectroscopic 
evidence of a companion star using the techniques described in Section 6.1 of \citet{Marcy:2014}. The detection threshold for any companion star 
depends on the RV separation between the primary star and
the supposed secondary star. For all RV separations greater than 20 
km/s, we would detect (at $3\sigma$) any
companions that are at least 2\% as bright (in the optical) as the primary star. 
For RV separations of 10 km/s, the detection threshold rises to 3\% as 
bright as the primary star, and for RV separations smaller than 10 km/s, 
the detection threshold rises rapidly to unity for FGK stars but remains 
at 3\% for M dwarfs (unless the primary itself is an M star) due to their very different spectra.   The poor 
detectability of FGK-type companion stars having little Doppler offset 
is due to the absorption lines that overlap.  Companions from 3500 -- 
6500 K would be detected by this technique, as such stars have strong 
absorption lines.

This spectroscopic technique only detects companions located angularly within a narrow rectangle of solid angle of size $0.87\arcsec \times 2\arcsec$,  centered of the primary stars, set by the width of the entrance slit to the spectrometer and the seeing of typically $1\arcsec$.    This small angular separation implies that most ``secondary" stars found by this technique would be gravitationally bound to the primary, and thus reside at the same distance from Earth.

The overwhelming majority of the 259 host stars of multis showed no spectral evidence of a stellar companion to the host star within $0.4\arcsec$ of the primary 
star, corresponding to half of the slit-width ($0.87\arcsec$) of the 
Keck-HIRES spectrometer.  We find clear evidence for a stellar companion only around KOI-2311. This companion appears to have a radial velocity relative to the primary star of 13 km/s and is 15\%$\pm$5\% as bright in the continuum as the primary star. However, this system fails our validation tests in multiple ways: The 
longer period candidate, KOI-2311.01, has S/N $< 10$ (fails criterion 
$iii$), so as KOI-2311 has only one other candidate, it fails to 
qualify as a multi for our statistical tests discussed in Sections 3 and 
4.  Neither planet candidate can be localized in the plane of the sky 
well enough to qualify for validation (criterion $vii$).  Furthermore, the 
transit of the shorter period candidate is very long for its orbital 
period, and would require a very eccentric orbit to be consistent with 
the stellar density estimated from spectroscopic classification; in 
contrast, the duration of the outer candidate's transit is very short 
for its orbital period.

Thus, of the 181 multis that we validate as multiple planet systems for 
which we have Keck spectra, \emph{none} show spectroscopic evidence for 
binarity. While wide binaries as well as binaries of extreme luminosity ratio would escape our detection technique, the  absence of identified spectroscopic companions to multis suggests that binary stars with separations of less than a few AU tend not to possess flat multiple planet systems on S-type orbits.   Kolbl \& Marcy (in preparation) will present a more detailed analysis of the implication of the paucity of spectroscopic binaries among multi-planet hosts.

\subsection{High-Resolution Imaging}

Many of the host stars of multis have been imaged at high resolution.  The \emph{coverage is highly nonuniform}, but when nearby stars were detected, we incorporated their presence into our centroid studies presented in 
 Paper III. Information about nearby stars is listed in Table 5 of Paper III.  Most of the  stars observed near putative multi hosts are $>$ 1$''$ from the target star, and could well be chance-alignment blends. For such targets,  we only validate candidates whose transit locations imply that they are on the target star rather than on the sky-plane neighbor. In two cases, neighbors are $\sim$ 1$''$ from the target star, and four targets have neighbors within 0.25$''$. None of the candidates of these six multis targets near neighbors have adequate centroid data to determine which of the two proximate stars is the source of planet candidates.  For two of the six, we have strong evidence that the proximate stars are physical binaries, and we validate their candidates as planets (without assigning them to a particular member of the binary) in Section 9, but we leave the candidates of stars with proximate neighbors that might not be bound companions as unvalidated.

\section{Binary Star Systems with Multiple Transiting  Planets}

Two targets have multiple candidates that  can all be validated as planets and assigned to a specific binary star system, but not assigned to a particular star in these stellar systems.  In the case of KOI-284, it is clear that the three planets do not all orbit the same star.  The five planets of KOI-1422 may represent one or two planetary systems. Sizes and periods of these planets are listed in Table \ref{tab:binaryplanets}.  We discuss the KOI-284 and KOI-1422 systems in more detail below.
  
\begin{deluxetable}{ccccc}
\tabletypesize{\tiny}%
\tablecolumns{5} \tablewidth{0pc} 
\tablecaption{Periods and Sizes of Validated Planets Known to be in Binaries}
\tablehead{ 
	\colhead{KOI number} &
	\colhead{Orbital Period (days)} &
	\colhead{Size if Orbiting Primary ($R_p/R_\star; R_\oplus$)} &
	\colhead{Size if Orbiting Secondary ($R_p/R_\star; R_\oplus$)} } 
\startdata
284.03 &6.17819 & $0.0126\pm0.0007$, $1.55\pm0.31$ & $0.0139\pm0.0007$, $1.62\pm0.37$ \\
284.02 &6.41491 & $0.0135\pm0.0007$, $1.67\pm0.34$ & $0.0148\pm0.0007$, $1.73\pm0.40$ \\
284.01 &18.01020 & $0.0177\pm0.0009$, $2.18\pm0.44$ & $0.0195\pm0.0009$, $2.28\pm0.52$ \\\hline
1422.03 &3.62146 & $0.0173 \pm 0.0007$, $1.13\pm0.06$ & $0.0353 \pm 0.0023$, $1.71\pm0.14$ \\
1422.01 &5.84165& $0.0412\pm0.0015$, $2.68\pm0.14$ & $0.0800\pm0.0052$, $3.89\pm0.31$\\
1422.02 &19.85024 & $0.0427\pm 0.0016$, $2.78\pm0.14$ & $0.0829 \pm 0.0054$, $4.03\pm0.32$\\
1422.05 &34.14235 & $0.0302 \pm 0.0011$, $1.97\pm0.10$ & $0.0587 \pm 0.0038$, $2.85\pm0.23$ \\
1422.04 &63.3358& $0.0355 \pm 0.0014$, $2.31\pm0.12$ & $0.0686 \pm 0.0043$, $3.33\pm0.26$ \\

\enddata
\label{tab:binaryplanets}
\end{deluxetable}

\subsection{Kepler-132 = KOI-284: A Binary With Both Stars Possessing Transiting  Planets}

KOI-284, a  bright (\ik magnitude $Kp  = 11.8$) target with three planetary candidates having periods of 6.18, 6.42, 18.0 days 
and nominal sizes of $\sim$ 1.4 R$_\oplus$ (Earth radii), is the only one of 170 multi-candidate system identified in \cite{Borucki:2011} 
that would be clearly dynamically unstable if all of the planets orbited the same star 
\citep{Lissauer:2011b}.  Speckle images of KOI-284 obtained on 24 June 2010 revealed two stars differing in 
brightness by less than one magnitude and separated from one another by slightly less than 1$''$ 
\citep{Howell:2011}.  Spectroscopic observations of each individual star obtained at Keck in 2011 show 
the two stars to have nearly identical spectra and have a difference in radial velocity of 0.94 $\pm$ 0.1 
km/s. The nearly identical velocities are consistent with their being gravitationally bound.
Their separation of $0.9''$ at a distance of roughly 500 pc implies a projected separation of 450 AU; the 
relative orbital velocities for two 1 $M_\odot$ stars on a circular orbit with semimajor axis $a$ = 450 AU 
is 2 km/s.  Analysis of the spectra together with Yonsei-Yale stellar models yields the following characteristics:  

\noindent Western component:
T$_{\rm eff}$:   $5963 \pm 100$~K;
$\log g$:   $4.36 \pm 0.15$;
Fe/H: $-0.26 \pm 0.1$;
$L_\star = 1.44^{+0.35}_{-0.64}~L_{\odot}$; $R_\star = 1.13 \pm 0.22~R_{\odot}$.

\noindent Eastern component:
T$_{\rm eff}$:    $5792 \pm 100$~K;
$\log g$:    $4.35 \pm 0.15$;
Fe/H:   $-0.30 \pm 0.1$;
$L_\star = 1.16^{+0.34}_{-0.70}~L_{\odot}$;
$R_\star = 1.07 \pm 0.24~R_{\odot}$.

\begin{figure}
\includegraphics[scale=0.8]{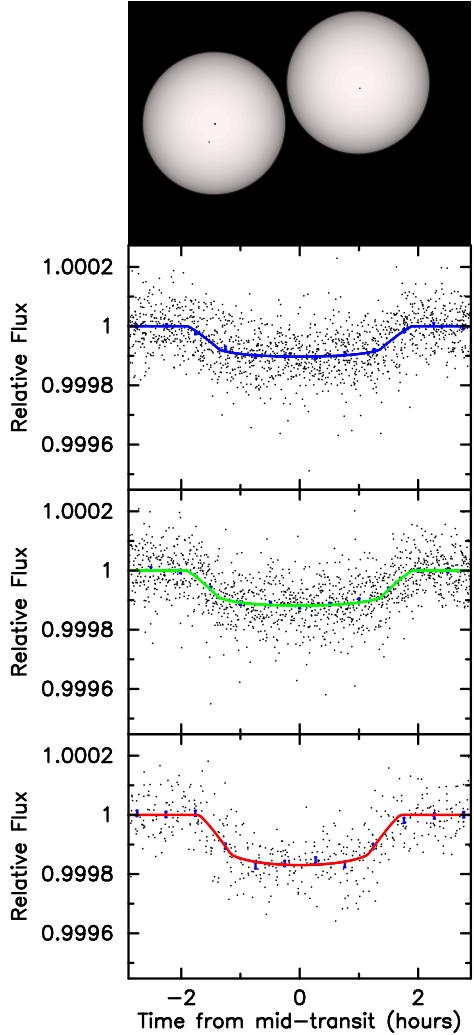} 
\caption{Transits of all three of the planets observed in the Kepler-132 = KOI-284 binary star system.  The beauty shot in the upper panel displays the sizes of the stars and planet candidates to a uniform scale. Planets and candidates are displayed with distance below the middle of the star corresponding to the transit impact parameter. The lower panels show the detrended \ik flux from the target phased at the period of each transit signal and zoomed 
to a region around mid-transit,  shown in order of increasing orbital period. Black dots represent individual \ik long cadence observations.  The blue bars are the data binned to 30 minutes in phase with 1 $\sigma$ uncertainties. The colored curves show the model transit fits, with colors corresponding to the last two digits of KOI designators as follows: red = .01, green = .02 and blue = .03. In each panel, the best-fit model for the other two planet candidates was 
removed before plotting. }
\label{fig:284planettransits}
\end{figure}

These observations suggest that the three candidates may well all be transiting planets in the 
same stellar system, with one of the stars hosting one of the six day period planets and the other two 
planets orbiting the other star in the binary. This configuration would be dynamically stable and thus passes our criterion ($ix$). All three candidates pass all of our other tests for validation as planets, so we validate them as planets and name the system Kepler-132.  We do not, however, assign letter designations to the individual planets.  We anticipate that eventually the two planets with periods between 6 and 7 days will be identified as orbiting different host stars and designated Kepler-132A b and Kepler-132B b, and the 18 day period planet will be designated either Kepler-132A c or Kepler-132B c depending on which star it orbits. Planetary radius estimates are increased by roughly a factor of $2^{1/2}$ to account for the dilution of light by the non-host star (the dilution is $45 \pm 5\%$ for the primary and $55 \pm 5\%$ for the secondary), and numerical values incorporating this correction are presented in Table \ref{tab:binaryplanets}. Figure \ref{fig:284planettransits} shows the sizes of the planets relative to the stars and folded light curves of the transits of each of the three planets. A detailed analysis of both \ik and ground-based data to characterize the fascinating Kepler-132 system is currently underway (Bryson et al., in preparation).

\subsection{Kepler-296 = KOI-1422: Two Small Stars With a Total of Five Transiting Planets}

KOI-1422 has  five planet candidates, all  of which pass our criteria for planetary validation with one caveat:
High resolution imaging observations of KOI-1422 with various telescopes
have shown this target to be a
tight binary with separation of about 0.22$''$.
Star \& Gilliland (in preparation) have performed an extensive analysis of the stars
relying primarily on  the F555W and F775W photometry from
HST/WFC3 data, which provide images in well calibrated optical filters
that
cleanly separate the two components.  Adopting the HST photometry
and an empirical transformation to the \ik bandpass shows that 80\% of 
the
light detected by \ik
comes from the brighter component, and 20\% from the fainter, with
an estimated uncertainty of 3\%.

\begin{figure}
\begin{center}
\includegraphics[scale=0.5]{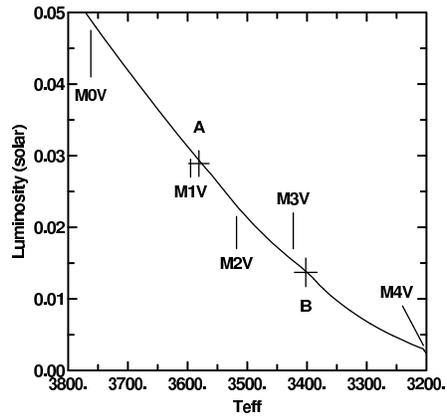}
\end{center}
\caption[]{HR diagram for KOI-1422, with the curve being the
  5 Gyr, solar abundance empirical isochrone.  Superposed with
  {\bf A} and {\bf B} labels are the inferred location following synthetic
  photometry using \cite{Pickles:1998} standard spectra.  Spectral
  types are adopted from \cite{Lepine:2013}.}
\label{fig:isochrone}
\end{figure}

Star and Gilliland (in preparation) performed isochrone matching
using the Dartmouth isochrones \citep{Dotter:2008, Feiden:2011}
to fit for the separate F555W-F775W colors of both components A and B,
the magnitude difference of the two components in F775W, and have
included $i-J$  for the combined A + B components since these are not resolved in the source photometry.
The two components fall
nicely on a single isochrone.  Given the close angular separation
and the ease of matching the components to a single isochrone,
Star and Gilliland  conclude that components A and B of KOI-1422 are
almost certainly
a bound, coeval system.  However, the temperatures are well
above the \cite{Muirhead:2012b} value based on $K$-band spectroscopy.
An alternate, empirical isochrone fit using standard spectra
and synthetic photometry yields a much cooler solution
consistent with \cite{Muirhead:2012b}.
Figure \ref{fig:isochrone} shows the empirical isochrone match, which is also excellent.

Thus, there is very strong evidence that KOI-1422 is a physical binary that possesses five transiting planets.   Ordering the planets by orbital period gives three of the period ratios of successive entries between 1.6 and 1.9 and the fourth is 3.4.  As none of these planets are giants, they could all be on stable orbits around the same star. Any subset of planets associated with either stellar component
   is also acceptable.

For both empirical and Dartmouth stellar isochrones, component B has a radius $\sim$0.75 that of A,
and is about two spectral sub-types later than component A.
Based on their two alternate isochrone matching exercises,
Star and Gilliland (in preparation) find that component A could have (radius, $T_{\rm
eff}$, spectral type)
= (0.596 $R_\sun$, 4045 K, K7V) (Dartmouth), or (0.445 $R_\sun$,
3581 K, M1V) (empirical).
Adopting the \cite{Muirhead:2012b} result to break our degeneracy,
we choose results from the empirical isochrone fitting and
use the empirical stellar parameters for our estimates of planetary sizes given in Table  \ref{tab:binaryplanets}.



\section{Planetary Systems with Five or More Candidate Transiting Planets}

\subsection{The Kepler-90 = KOI-351 Seven-Planet System}

 The  KOI-351 system was initially announced to host three planetary
candidates by \cite{Borucki:2011}. We found an additional four candidates
 with the quasi-periodic automated transit search algorithm
\citep{Carter:2013}; three of these additional planets were subsequently
identified by the Q1-Q8 \ik pipeline and are thus included in Table 3 of Paper III.   In total, as presented by one of us (E.~A.) at the Exoplanets conference hosted by the Aspen Center for Physics in February 2013, this
star hosts seven transiting planet candidates with periods ranging
from 7 days to 330 days\footnote{Two other groups announced independent discoveries of KOI-351.07 later in 2013. \cite{Schmitt:2013} reported their discovery
as a system with seven planet candidates. \cite{Cabrera:2013} called all of the candidates planets using plausibility
arguments based upon a rough consistency of transit durations with those of a system of planets on circular orbits
around the same star. Such relationships do provide evidence supporting the planetary system hypothesis, but are
inadequate to validate planets at the level of confidence required by our study.}. 
The stellar density was constrained from a joint fit to the transit durations and impact parameters, assuming the eccentricities were constrained such that the apoapse and periapse of adjacent orbits were ordered by orbital period.  A Markov chain was used to find a density of $\rho_\star = 0.977\pm0.114$ g~cm$^{-3}$. The stellar parameters were derived following the procedures of Torres et al. (2008) from a fit of Yonsei-Yale isochrones (Yi et al. 2001) to the spectroscopically measured temperature (5994 $\pm$ 77 K) and metallicity ([Fe/H] = +0.04 $\pm$ 0.10) of the star, and using the stellar density from the light curve fit.  This procedure yielded the values $M_\star = 1.118_{-0.058}^{+0.039}~M_\odot$ and $R_\star = 1.166_{-0.044}^{+0.059}~R_\odot$. 

\begin{figure}
\includegraphics[scale=0.8]{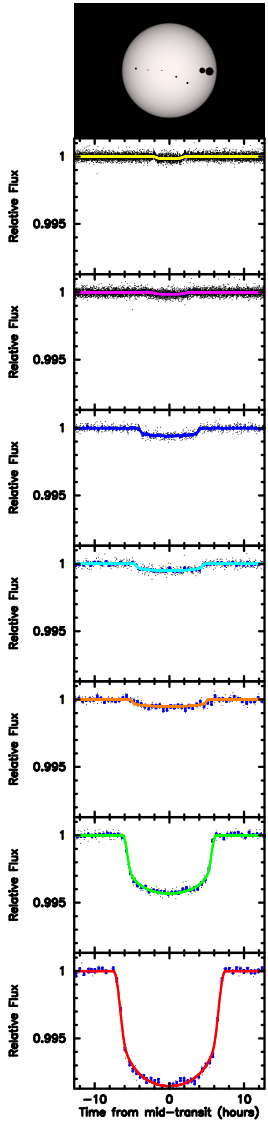}
\caption{Transits of all of the planets in the KOI-351 7-planet system.  The beauty shot in the upper panel displays the sizes of the star and planet candidates to a uniform scale. The color of the star and the impact parameters of the planetary transits reflect estimates of stellar and transit characteristics given in Section 10.1 and Table 6. Planets are shown black dots with sizes scaled appropriately relative to the star. Planets are displayed with distance below the middle of the star corresponding to the transit impact parameter. The lower panels show the detrended \ik flux from the host star during Q1 -- Q10 phased at the period of each transit signal and zoomed 
to a region around mid-transit,  shown in order of increasing orbital period. Black dots represent individual \ik long cadence observations.  The blue bars are the data binned to 30 minutes in phase with 1 $\sigma$ uncertainties. The colored curves show the model transit fits, with colors corresponding to the last two digits of KOI designators as follows: red = .01, green = .02, blue = .03, cyan = .04,  purple = .05, yellow = .06 and orange = .07. In each panel, the best-fit model for the other six planets was 
removed before plotting.  All panels have an identical vertical scale, to show the relative depths, and 
identical horizontal scale, to show the relative durations.}
\label{fig:7planettransits}
\end{figure}

We fit the light curves of all 7 planet candidates simultaneously with a standard quadratic limb-darkening transit light curve model. The inner two planets were assumed to have a periodic ephemeris, while for the outer five we allowed the individual times of transit to vary.  The characteristics of KOI-351's seven (candidate) planets are given in Table~\ref{tab:351planets} and transit light curves for all of them are shown in Figure~\ref{fig:7planettransits}.

\begin{deluxetable}{cccccccc}
\tabletypesize{\tiny}%
\tablecolumns{8} \tablewidth{0pc} 
\tablecaption{Kepler-90's Planetary System}

\tablehead{ 
	\colhead{Planet} &
	\colhead{Kepler-90 b} &
	\colhead{Kepler-90 c} &
	\colhead{Kepler-90 d} &
	\colhead{Kepler-90 e} &
	\colhead{Kepler-90 f} &
	\colhead{Kepler-90 g} &
	\colhead{Kepler-90 h} }
\startdata

%
KOI \# & KOI-351.06 &  KOI-351.05 &  KOI-351.03&  KOI-351.04 &  KOI-351.07 & KOI-351.02 & KOI-351.01 \cr
\hline
$T_0$  &  70.6797$\pm$ &  72.5208$\pm$ &  91.9622$\pm$ &  67.2952$\pm$ &  62.791 $\pm$   &  79.8448$\pm$ &  73.4992$\pm$ \cr
(BJD-2454900) &  0.0012       &   0.0038      & 0.0035       &  0.0079       &  0.011          & 0.0015       &   0.00085 \cr
$P$      & 7.008214$\pm$ & 8.718397$\pm$ & 59.73700$\pm$ & 91.94080$\pm$ & 124.9134 $\pm$  & 210.70287$\pm$ &  331.59940$\pm$   \cr
[days] & 0.000102      & 0.000324      & 0.00027 & 0.00078       &   0.0013        & 0.00037 &    0.00032\cr
duration [hr] &   3.72 $\pm$ 0.02 &   4.02 $\pm$ 0.02 &   7.90 $\pm$ 0.06 & 
  9.16 $\pm$ 0.08 &  10.03 $\pm$ 0.11 &  11.38 $\pm$ 0.06 &  13.21 $\pm$ 0.04\cr
depth [\%] & 0.0135 $\pm$ 0.0009 & 0.0175 $\pm$ 0.0009 & 0.0580 $\pm$ 0.0017 & 
0.0502 $\pm$ 0.0019 & 0.0642 $\pm$ 0.0032 & 0.4225 $\pm$ 0.0028 & 0.8246 $\pm$ 
0.0055 \cr
$R_p/R_\star$ & 0.0108 $\pm$ 0.0004 & 0.0122 $\pm$ 0.0003 & 0.0223 $\pm$ 0.0004 & 
0.0208 $\pm$ 0.0004 & 0.0234 $\pm$ 0.0006 & 0.0605 $\pm$ 0.0005 & 0.0840 $\pm$ 
0.0005   \cr
$b$ &  0.02 $\pm$  0.28 & -0.00 $\pm$  0.26 &  0.22 $\pm$  0.13 &  0.01 $\pm$ 
 0.28 & -0.00 $\pm$  0.25  & -0.01 $\pm$  0.24  & -0.04 $\pm$  0.23 \cr 
$R_p$ [R$_{\oplus}$] &  1.37 $\pm$ 0.08 &  1.55 $\pm$ 0.09 &  2.83 $\pm$ 0.15 & 
 2.64 $\pm$ 0.14 &  2.98 $\pm$ 0.17  &  7.65 $\pm$ 0.38  & 10.69 $\pm$ 0.53 \cr
 $S/S_\odot$ &  292.51 $\pm$ 23.27 &  217.55 $\pm$ 17.23 &   17.18 $\pm$  1.58
 &    9.61 $\pm$  0.82 &    6.31 $\pm$  0.52  &    3.17 $\pm$  0.26  &    1.74
 $\pm$  0.15 \cr
\enddata
\label{tab:351planets}
\end{deluxetable}

The inner pair of KOI-351's seven (candidate) planets are a little larger than Earth; the middle three are $\sim 2.6 - 3$ $R_\oplus$; the sixth is a sub-Saturn and the outermost a little larger than Saturn. Transit durations increase monotonically with orbital period and are consistent with near-central transits of a system of planets on low eccentricity orbits about the same star. All six of the planet candidates of KOI-351 that were found by the pipeline pass our standard validation criteria, so we validate them as planets.  The other candidate, KOI-351.07, passes all of our validation criteria apart from having been identified in the sample used for statistical studies, implying that it is a very strong planet candidate.  However, we require additional evidence supporting 351.07 in order to verify that it is a planet.

\begin{figure}
\begin{center}
\includegraphics[scale=0.7]{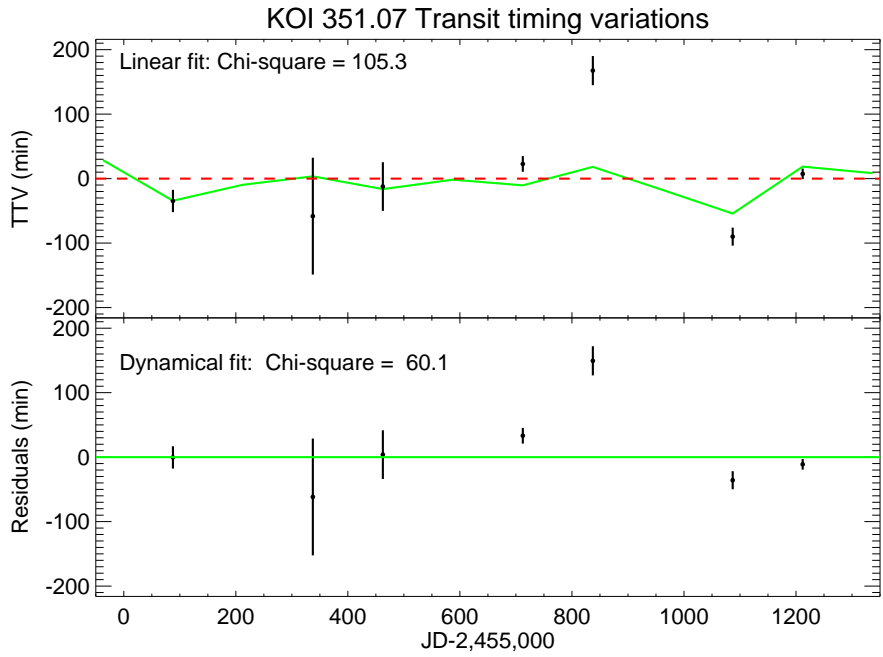} 
\end{center}
\caption{The transit times of (candidate) planet KOI-351.07 (black points with error bars), minus computed models with: (top) a single planet on a periodic orbit;  (bottom) a planet on a circular orbit that is perturbed by the two outer planets whose parameters are fixed, as described in the text. The transit times predicted by the perturbed model are represented by the solid green curve in each panel.}
\label{fig:koi351.07TTs}
\end{figure}

Transit timing variations provide dynamical evidence that KOI-351.07 is a planet in the same system as the validated planets.  A complete dynamical analysis of the seven-planet KOI-351 system is beyond the scope of this work.  But  the outer two planets are much larger than the inner five, and a dynamical fit to the observed transit times of this pair, neglecting the other planets, models their  TTVs very well. We then fit the observed transit times for the candidate 351.07 by allowing the period and epoch of this candidate to vary, but fixing the orbit to be circular at epoch, giving two free parameters. We performed two such fits, in one case assuming that 351.07 was on a Keplerian orbit, whereas in  the other, it was perturbed by the two large planets in the system using parameters for the large planets fixed to values computed as described above.  Not surprisingly, given that the models do not allow for 351.07 to be on an eccentric orbit at epoch and do not account for perturbations of the planets interior to its orbit, neither fit is formally adequate. Nonetheless, as shown in Figure~\ref{fig:koi351.07TTs}, the fit with the two outer planets was much better, reducing $\chi^2$ from 105 to 60.  This result provides strong additional support to the claim that KOI-351.07 is a planet and orbits the same star as the two largest transiting planets in the system. Therefore, we validate KOI-351 as a 7-planet system.  The parameters of all seven planets are given in Table~\ref{tab:351planets}. A future paper (Agol et al., in preparation) will present a more detailed
analysis of this system.
 
\subsection{Targets with Six Planet Candidates}
\begin{figure}
\includegraphics[scale=0.8]{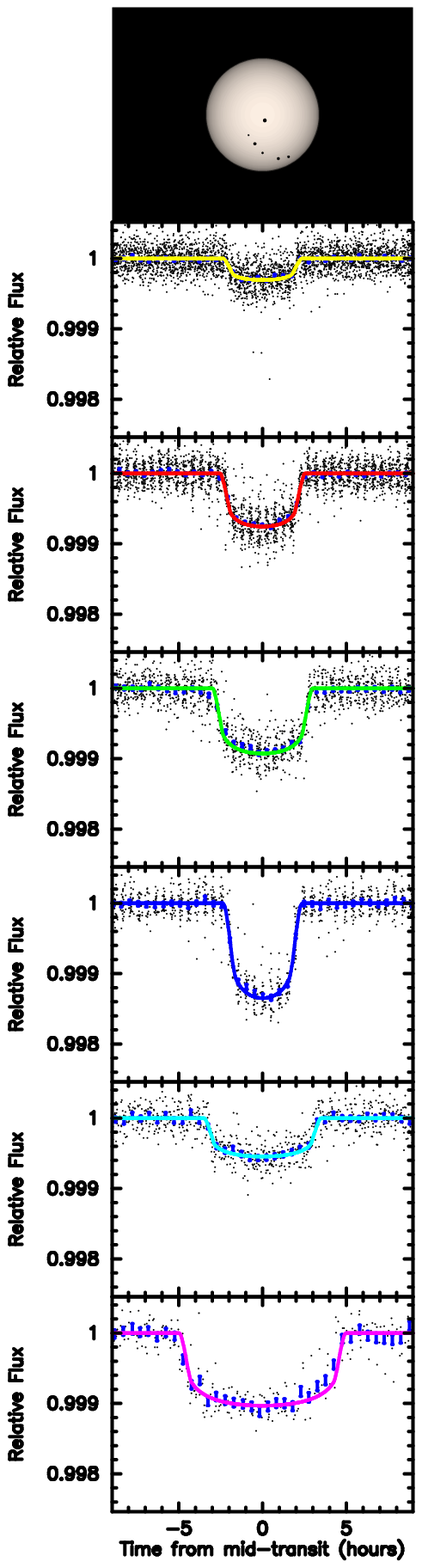} 
\includegraphics[scale=0.8]{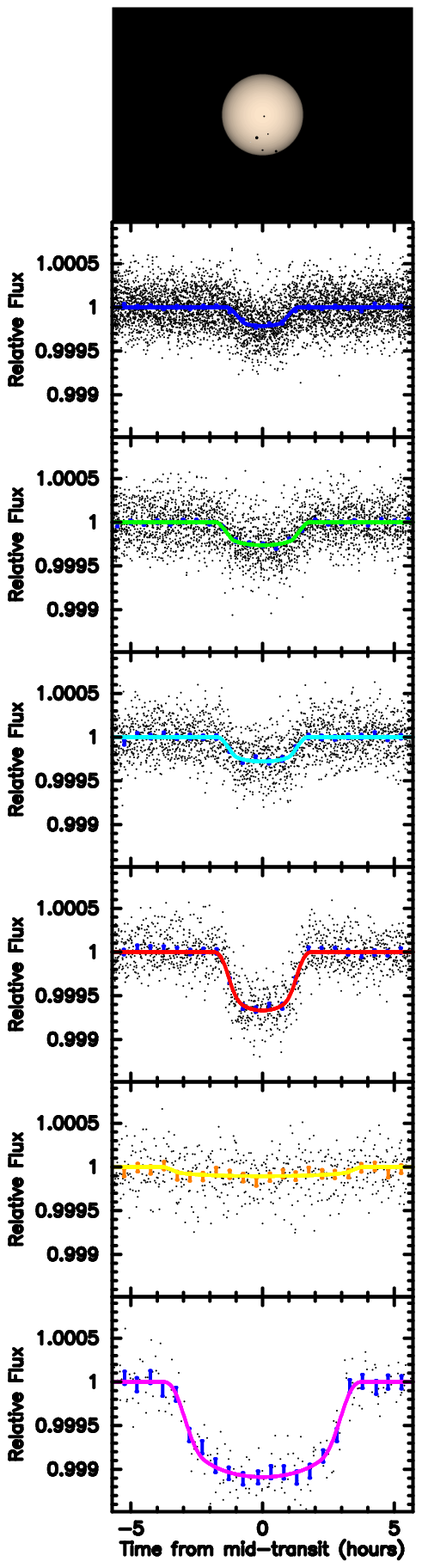} 
\caption{Transits of all of the planets/candidates in both systems with six planet candidates based upon parameters given in Paper III. See caption of Figure \ref{fig:7planettransits} for a detailed explanation.  The beauty shots in the upper panels display the sizes of the stars and planet candidates to a uniform scale. The unverified candidate KOI-505.05 is distinguished from the verified planets by appearing in green in the beauty shot and additionally its individual data points appear gray rather than black and binned data are shown in orange rather than blue. The left panel shows Kepler-11 =  KOI-157 and the right panel represents Kepler-169 = KOI-505; note that the time and depth scales  are uniform within individual systems but differ from system to system.}
\label{fig:6planettransits}
\end{figure}

 Two targets possess six transiting planet candidates.  Transit light curves for all of these candidates are shown in Figure \ref{fig:6planettransits}. All six candidates in KOI-157 pass our validation tests, but as these candidates have already been verified as Kepler-11 b-g by \cite{Lissauer:2011a} and  analyzed more extensively by  \cite{Lissauer:2013}, we do not consider them further herein.

Although   KOI-505 has six planet candidates, all identified in the Q1-Q8 transit searches, one of these candidates is a marginal detection with S/N = 7.1. 
All five of KOI-505's other planet candidates meet all of our  validation criteria, so we validate these five as planets and name the system Kepler-169. 
Close-ups of transits of all five planets and that of the unvalidated candidate are shown in Figure \ref{fig:6planettransits}.  
 No other star is evident in the stellar spectrum or in high-resolution images.  All five planets have radii intermediate between those of Earth and Neptune.  Orbital periods of the inner four planets range from 3.25 days to 13.8 days, with the fifth planet much farther out and having a period of 87.1 days. Planets c and d have the smallest period ratio, 1.347, placing them a bit wide of the 4:3 mean motion resonance (MMR); other planet pairs are far from any strong resonances. The period of the unvalidated and highly suspect candidate is 29.9 days.

\subsection{Other Five-Planet Systems}

\begin{figure}
\includegraphics[scale=0.6]{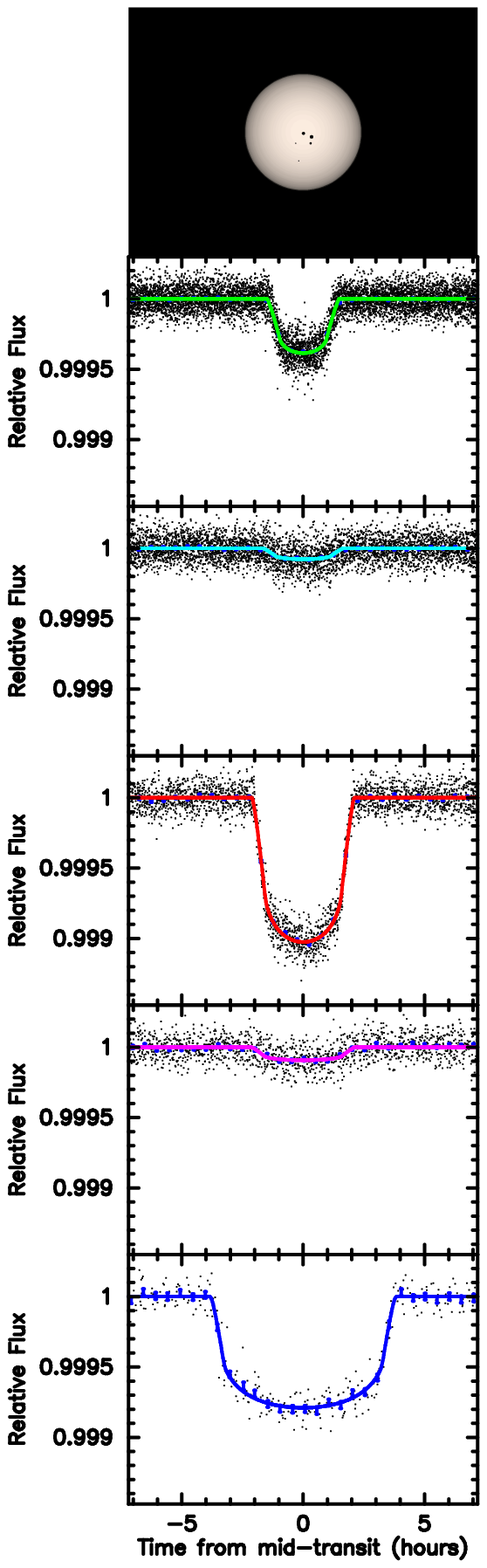}  
\includegraphics[scale=0.6]{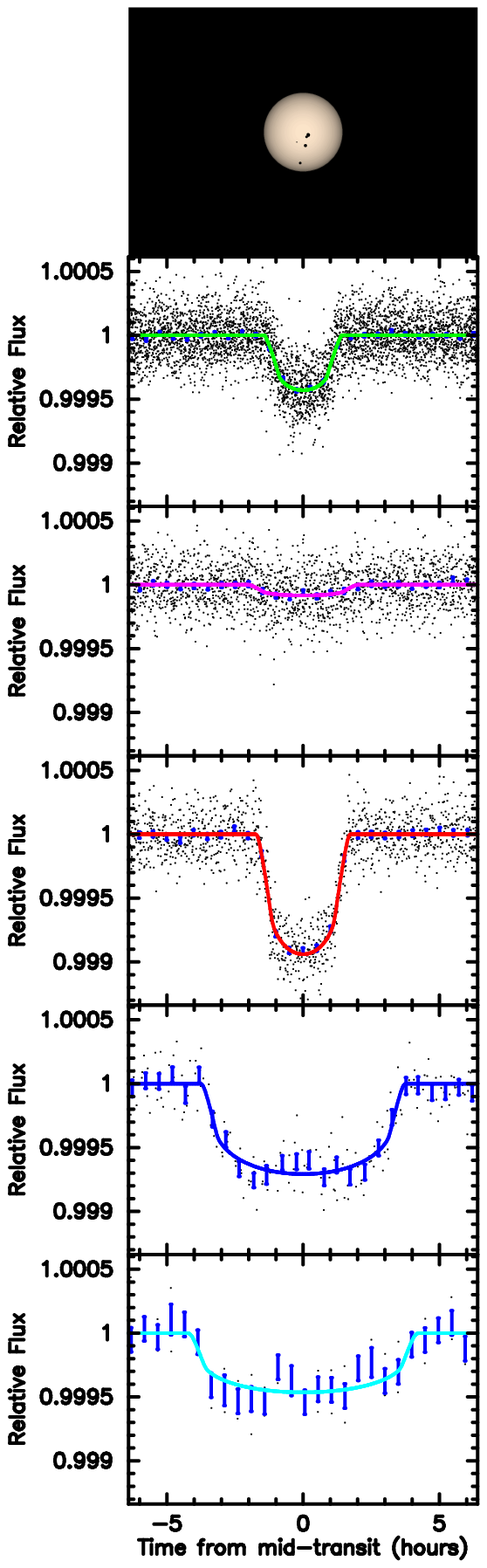}  
\includegraphics[scale=0.6]{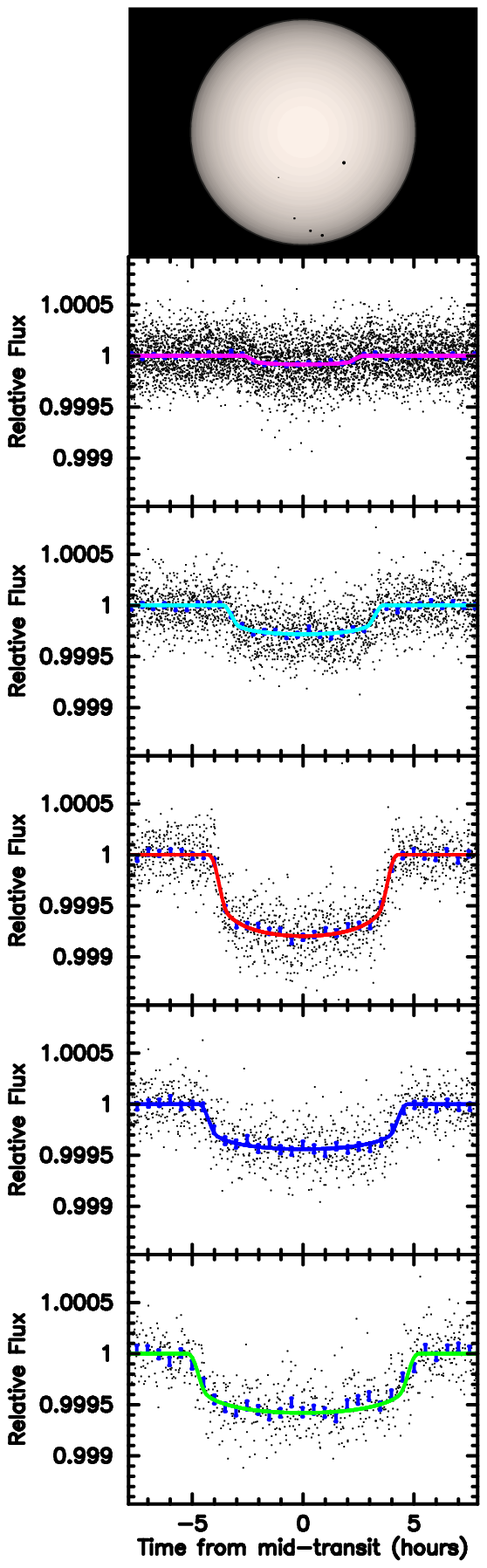}  
\includegraphics[scale=0.6]{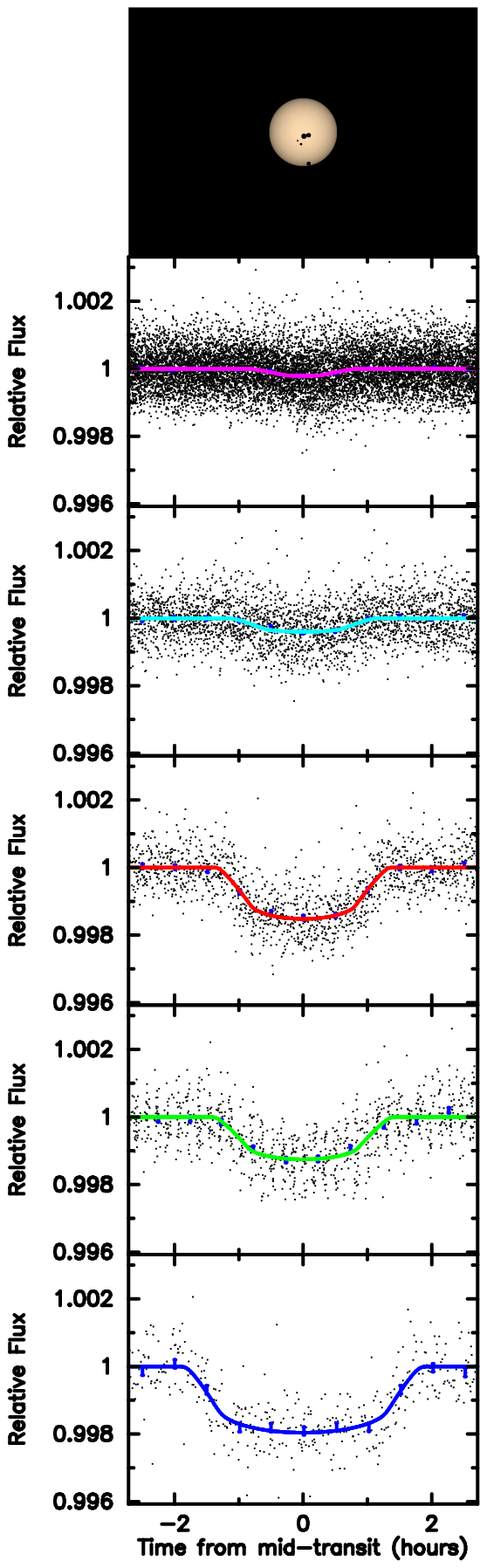}  
\caption{Transits of all of the planets in previously-verified 5-planet systems.  See the caption to Figure \ref{fig:7planettransits} for an explanation of the figure format. The successive panels (left to right) show KOIs 70 (= Kepler-20), 701 (= Kepler-62), 707 (= Kepler-33) and 952 (= Kepler-32). Note that the time and depth scales  are uniform within individual systems but differ from system to system.}
\label{fig:old5}
\end{figure}

The four previously-verified \ik five planet systems are illustrated in Figure \ref{fig:old5}.  All of the planets in KOI-701 = Kepler-62 \citep{Borucki:2013} and KOI-707 = Kepler-33 (Paper I) pass all of our validation tests.  The largest three planets orbiting KOI-70 = Kepler-20 \citep{Gautier:2012} also passed all of our tests, but we could not localize transit signals of the two Earth-size planets in that system \citep{Fressin:2012} to revalidate them via our techniques. The other previously-announced  5-planet system, KOI-952 = Kepler-32,  has four planets meeting our validation criteria, with the fifth having too short an orbital period for us to consider.   

The fact that some previously verified \ik planets were not revalidated by our techniques should not cast doubts on their earlier validations/confirmations.  Our techniques have been developed to validate large numbers of planets.  Blender pays a great deal of individual attention to each planet candidate that it considers using very careful fitting of the light curve shape, etc., and thus in some cases the centroid localization required for validation via Blender is not as strict as the criteria employed in our analysis.  Moreover, we paid less attention to those candidates that had already been validated as planets.  

\begin{figure}
\includegraphics[scale=0.5]{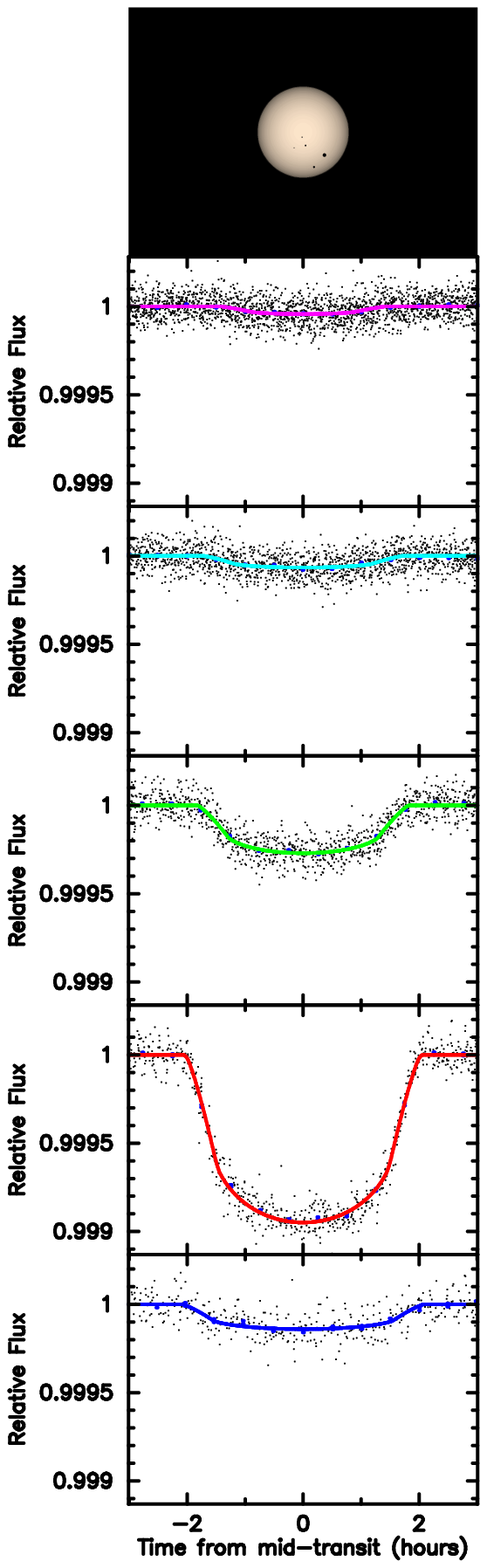} 
\includegraphics[scale=0.5]{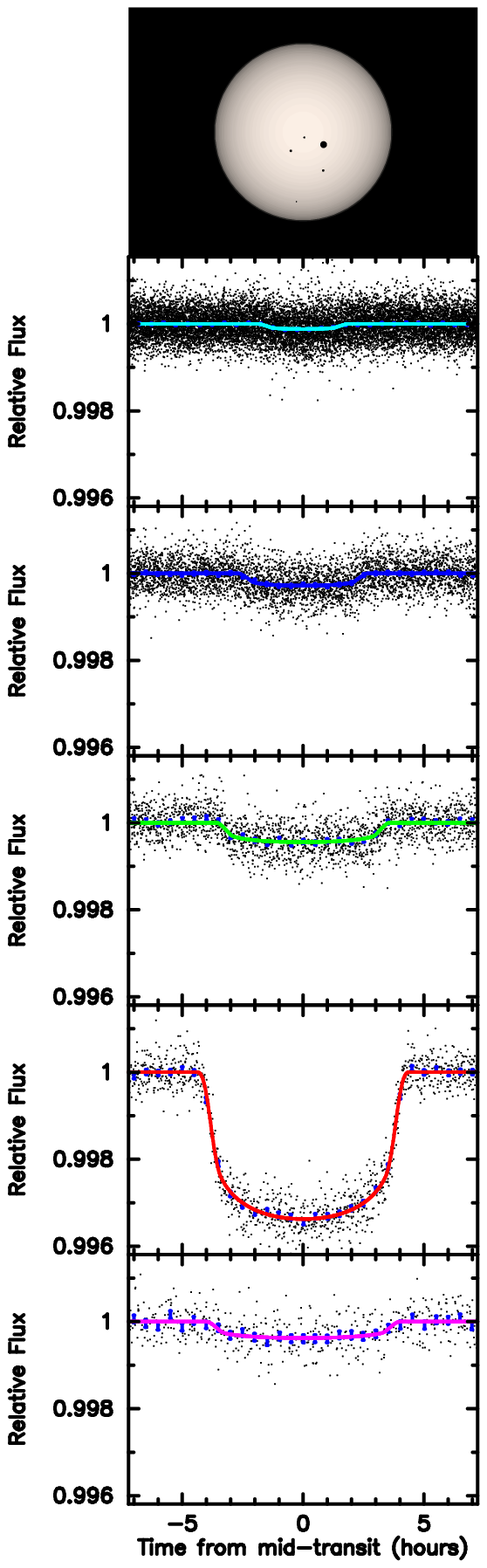} 
\includegraphics[scale=0.5]{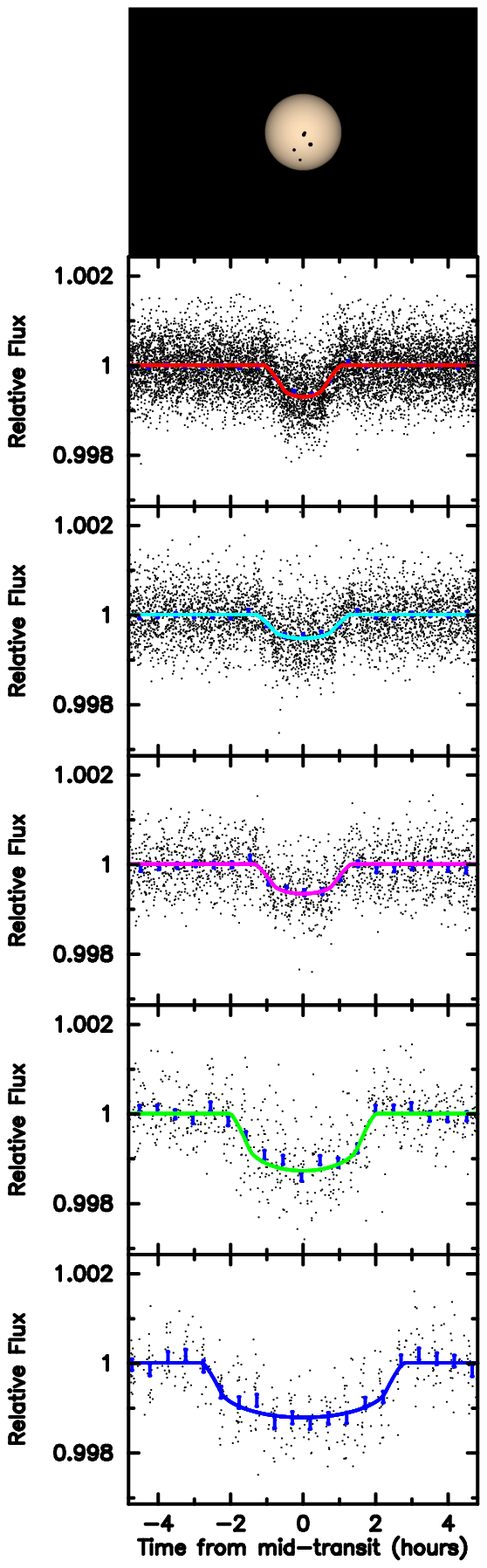} 
\includegraphics[scale=0.5]{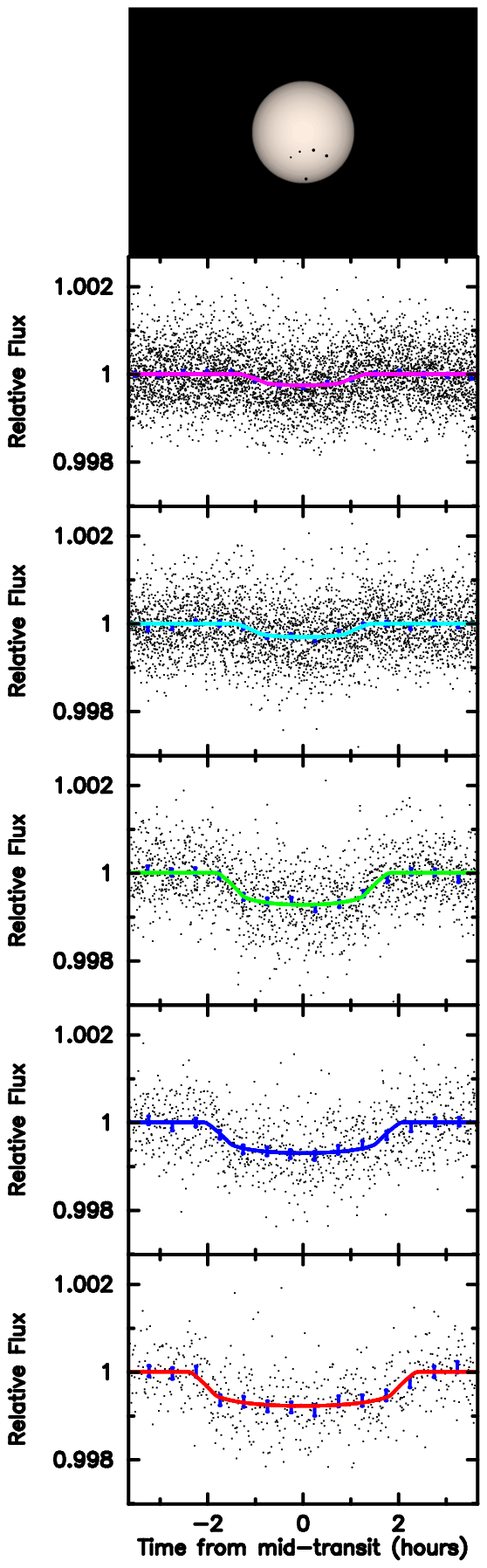} 
\includegraphics[scale=0.5]{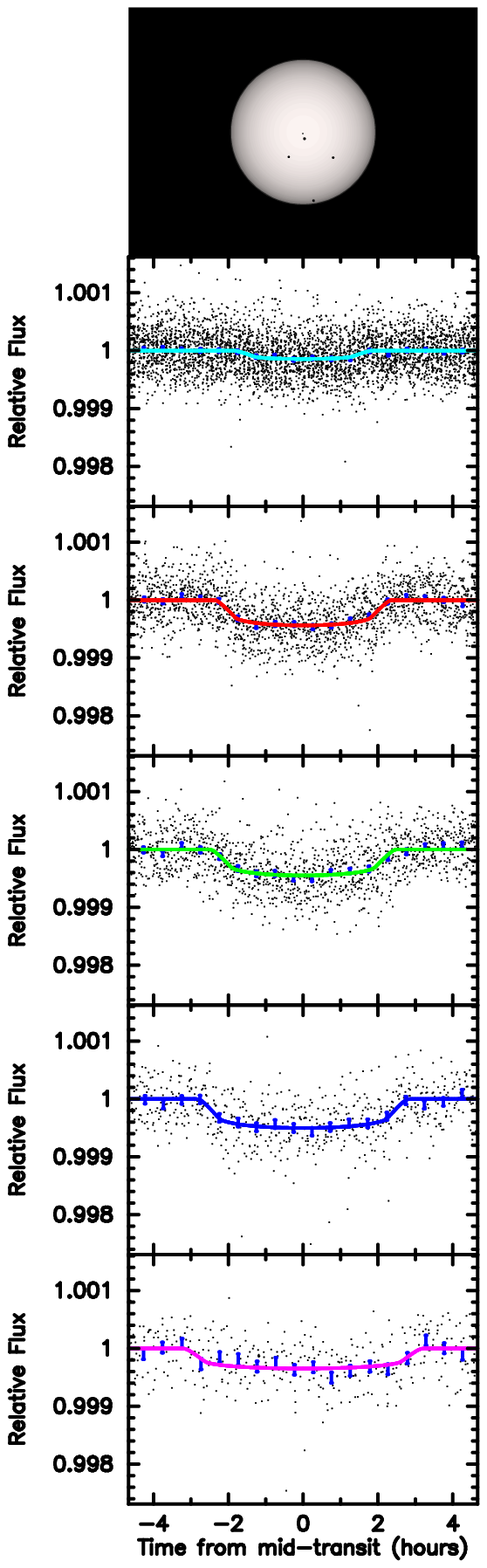} 
\caption{Transits of all of the planets in  5-planet systems that include newly-validated planets. We omit KOI-1422 from this diagram because the planets validated in this binary star system may not all orbit the same stellar component. See the caption to Figure \ref{fig:7planettransits} for an explanation of the figure format. The successive panels (left to right) show KOIs 82 (= Kepler-102),  834 (= Kepler-238), 904 (= Kepler-55), 1364 (= Kepler-292) and 1589 (= Kepler-84). Note that the time and depth scales  are uniform within individual systems but differ from system to system.}
\label{fig:good5}
\end{figure}

Figure \ref{fig:good5} illustrates the five other 5-planet systems validated herein. 
 The outer two planets of one of these systems, KOI-904 = Kepler-55, which lie near the 3:2 MMR and have sizes $\gtrsim 2~R_\oplus$, were previously confirmed via TTVs \citep{Steffen:2013a}.  This star's three inner planets all have radii $\sim 1.5~R_\oplus$ and period ratios from 2.08 -- 2.74. The radius of the inner planet of Kepler-84 = KOI-1589 is $\sim 1.4~R_\oplus$, whereas the outer four planets are roughly twice this size. Period ratios range from 1.48 to 2.13, with the tightest pairing being the only one near a strong MMR and consisting of the two planets previously confirmed using TTVs \citep{Xie:2013}.

Kepler-102 = KOI-82's five planets have periods between 5 and 28 days. Three of the five planets are smaller than Earth, one is slightly larger, and one just over twice the radius of Earth; the inner four planets are ordered by increasing size with increasing distance from the planet.  Period ratios of neighboring planets range from 1.338 to 1.70. The closely-spaced inner pair of planets, just wide of the 4:3 resonance, are both Mars-sized. No other planet pairs lie in or near first-order MMRs.  Despite the close spacing of the planets in period ratios, the small planet sizes suggest that this system is not as  closely-packed in the dynamical sense of lying near instability as are some of the other high-multiplicity systems discovered by \ikt. 

Kepler-238 = KOI-834 has five planets, four of which are in the $\sim 2 - 3~R_\oplus$ size range and the fifth about three times as large. These planets are more widely spaced in period ratio, with ratios of neighbors from 1.79 -- 2.94.  This system lacks significant resonances.

In the Kepler-292 = KOI-1364 system, the five detected planets are ordered by size, with the innermost a bit larger than Earth and the outermost about the size of Neptune. Adjacent planets have period ratios of 1.44 -- 1.90, so the system is fairly closely packed.  However, no planet pairs are in or close to strong MMRs.

\subsection{Incompletely Validated Systems with Five Planet Candidates}

\begin{figure}
\includegraphics[scale=0.6]{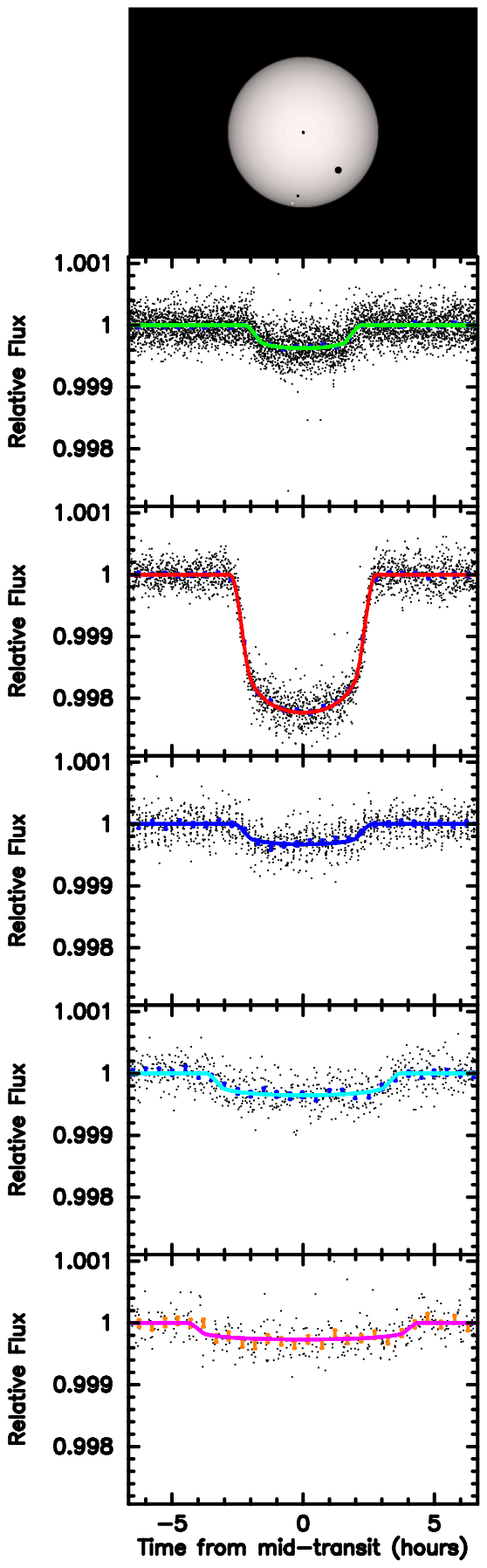} 
\includegraphics[scale=0.6]{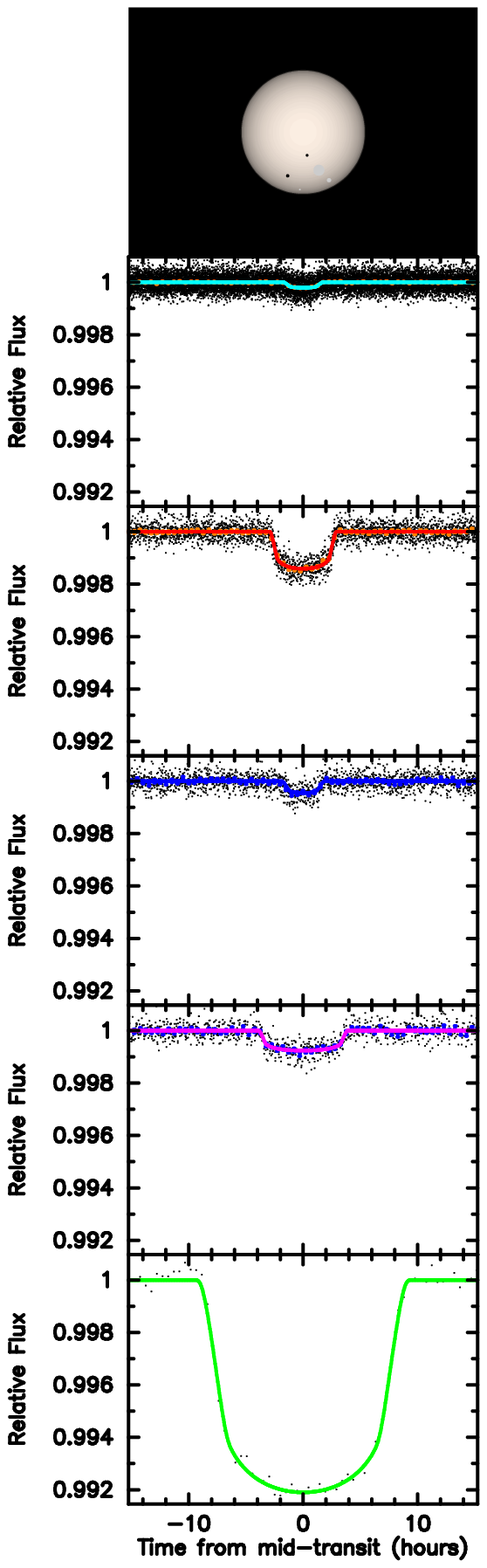} 
\includegraphics[scale=0.6]{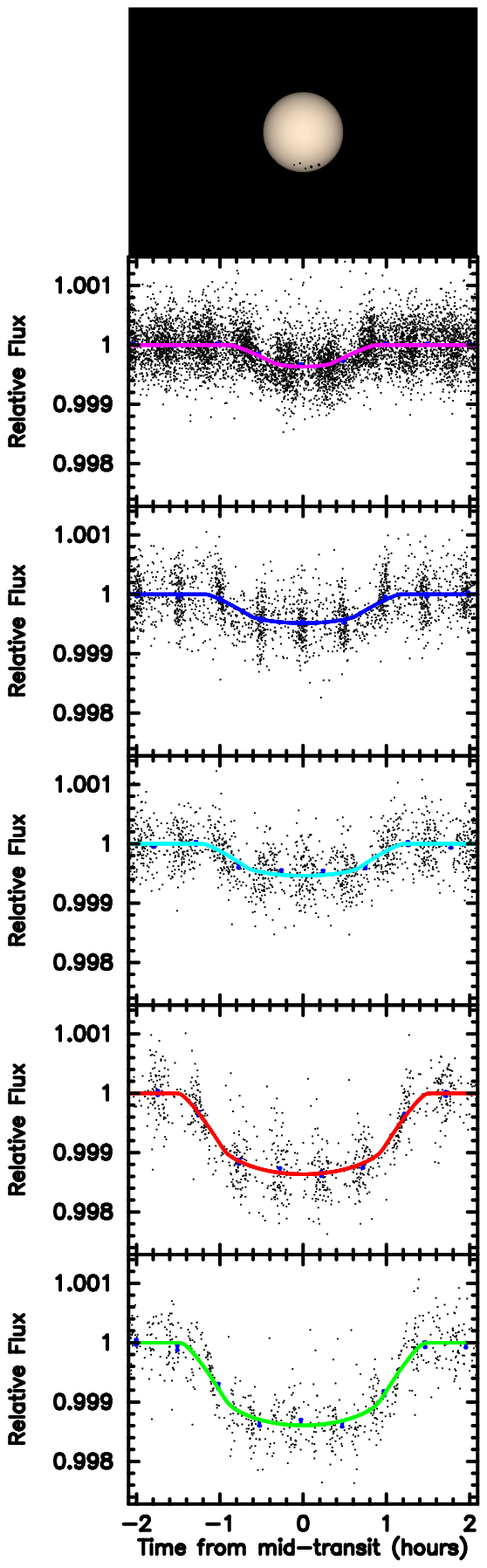} 
\caption{Transits of all of the planets  in systems with 5 planet candidates not all of which have yet been verified as planets. Verified planets are shown in black while other candidates are green.  See the captions to Figures \ref{fig:7planettransits} and \ref{fig:6planettransits} for an explanation of the figure format. The panels show KOIs 232 (Kepler-122, left), 435 (Kepler-154, center) and 500 (Kepler-80, right). The clustering of the data points for the four verified (resonant) planets of Kepler-80 is a stroboscopic pattern produced by  the phasing of the cadence and orbital timescales. }
\label{fig:mixed5}
\end{figure}

Figure \ref{fig:mixed5} illustrates the three five-candidate systems in which not all candidates have yet been verified to be planets.  The  outer candidate of KOI-232 = Kepler-122 barely misses our S/N cutoff for validation, but passes all of our  other requirements; thus it remains a strong candidate. One of the planets is $\sim 6~R_\oplus$ in size, while the three other planets that we validate as well as the unverified candidate are $\sim 2 - 3~R_\oplus$.  Period ratios among the neighboring validated planets are 1.76 -- 2.16; none are particularly close to the 2:1 MMR.  The candidate planet's period is 1.48 times as large as that of the outermost validated planet.   

KOI-500 = Kepler-80 has five planet candidates, all with orbital periods of under ten days.  We validate the four outer candidates, two of which were previously confirmed via TTVs \citep{Xie:2013}, to be planets; the inner candidate has too short an orbital period for us to validate, but passes all of our other validation criteria.  The four validated planets are near first-order mean motion resonances, but more interestingly the inner and outer threesomes are each in three-body resonances analogous to the Laplace resonance of the inner three Galilean satellites of Jupiter \citep{Lissauer:2011b}.  A detailed analysis of this fascinating system is underway (Ragozzine et al., in preparation).  

We are only able to validate two of Kepler-154 = KOI-435's five candidates as planets.  The validated planets and remaining candidates are widely spaced in period and far from resonances. 

\section{Kepler-223: A Four-Planet System in a Chain of Mean Motion Resonances}
The four candidates of KOI-730 were identified by \cite{Borucki:2011} and highlighted as being in a highly unusual chain of first-order resonances by \cite{Lissauer:2011b}.  The period ratios of consecutive candidates are now estimated (from inside outwards) as: 1.3337, 1.5017  and 1.3336, with uncertainties of $< 10^{-4}$, placing the pairs within $\sim 1$ part in 10$^3$ of exact mean motion resonances, but not exactly in MMR.  \cite{Lissauer:2011b} recognized that this set of period ratios was highly unlikely to have occurred by chance. Fewer than 1\% of periods drawn from a uniform distribution in a region that goes from just inside the 2:1 resonance to just outside the 6:5 resonance would lie this close to any of the four first-order mean motion commensurabilities seen in \ik planet candidates \citep{Fabrycky:2013}. Thus, these candidates were known to very likely be real planets without factoring in any general probability boost from multiplicity.  Independent of their resonant relationship, all four candidates satisfy all of our criteria for validation  and thereby are validated as planets; we name this system Kepler-223. 

\begin{figure}
\includegraphics[scale=0.6]{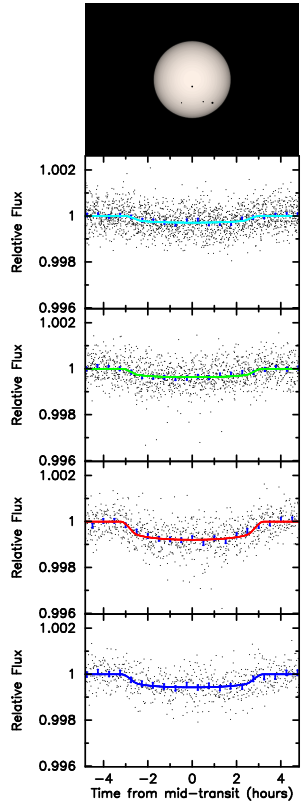}
\caption{Transits of all four planets  in the resonant system KOI-730 = Kepler-223.  See the caption to Figure \ref{fig:7planettransits} for an explanation of the figure format. }
\label{fig:ressys}
\end{figure}

The resonant relationships of the planets associated with KOI-730 = Kepler-223  imply that the likelihood that  all of them orbit the same star is even higher than that for the other planets that we are validating in this project.
Figure \ref{fig:ressys} illustrates the resonant  Kepler-223 system. 

\section{Conclusions}

A sizable majority of the planet candidates in \ik multis are validated as planets herein and in Paper III.  Statistical validation of \ik multis leads to a high level of confidence of planet status for each of the validated objects.   The vast majority of the objects passing our tests are planets orbiting the \ik target; combined with a much smaller group of planets orbiting bound stellar companions to the \ik target stars, we expect that  more than $99\%$ of the objects that we validate  herein and in Paper III are planets that travel around the Galaxy together with the star that we associate with them. 

Table \ref{tab:fpest} provides estimates of the number of FPs of various types 
expected among the 1033 members of the multiple planet candidate population that satisfy the six criteria listed in Section 3 and thus are considered for validation in this study. Under the 
pessimistic assumption that 10\% of the \ik single planet candidates are FPs, 
the total expected number of FPs is a mere 2.1. The most likely type of FP among the multiple planet systems that we have validated is a nominal two-planet system that actually contains 1 FP and 1 planet; a very small number of nominal two-planet systems may actually be 2 FPs.  The most likely pollutants of systems of higher multiplicity are 1 FP plus two or more planets; it is unlikely that any of the  systems of 3 or more validated planets contains more than one FP, although that possibility cannot be strictly excluded. 

Provided that one does 
not consider planets transiting a stellar companion to the \ik target as 
FPs, the \cite{Fressin:2013} and \cite{Santerne:2013} studies suggest an FP rate for singles of only 7\%, even after the upward adjustment of their estimates that 
neglect planetary multiplicity to apply to the subpopulation of singles 
as described in Appendix C.   Using an FP rate for singles of 7\% rather than 10\% reduces 
our estimate of the number of FPs in multis by more than 30\%.  These estimates are for the entire population of currently viable (candidate) multis that satisfy the six criteria listed in Section 3; the subset of multis validated herein and in Paper III is smaller and so the equivalent estimates are also smaller. The bottom line is that unless orbital planes of planetary systems are highly correlated with those of close binaries that they are bound to within triple (and higher multiplicity) star systems, $\gtrsim 99.8\%$ of the objects that we validate should be true transiting exoplanets. 

Although our work provides a catalog consisting of a large number of new, highly-reliable, planetary systems, it is far from complete, and it is biased in favor of common characteristics that makes systems amenable to our methods of statistical validation.  \emph{If a planet candidate has not been validated, that does not imply that it isn't a real planet; indeed, we expect that over 90\% of \ik multis not validated by our studies (nor identified to date as FPs) are nonetheless actual planets and planetary systems.}  Thus, while our sample is useful for those who want to study characteristics of individual systems without worry of contamination by false positives, the entire set of \ik candidate multi-planet systems is less biased and still highly reliable, making it superior for use in statistical studies of the distribution of planetary system characteristics.

\acknowledgments 
{\it Kepler} was competitively selected as the tenth Discovery mission. Funding for this mission is 
provided by NASA's Science Mission Directorate. We dedicate this paper to the memory of \ik Deputy Principal Investigator David G. Koch (1945 August 6 -- 2012 September 12), whose work over more than two decades was key  
to making the \ik mission a success. We acknowledge Ruth Murray-Clay for stimulating and valuable conversations about FP modes. Tony Dobrovolskis, Mike Haas, Billy Quarles,   Leslie Rogers and Alexandre Santerne provided constructive comments on the manuscript.
Elisabeth Adams assisted with the preparation of Figure \ref{fig:gallat}.

\section{ Appendix A: Galactic Latitude Distribution of \ik Targets and Planet Candidates}

\begin{figure}
\includegraphics[scale=0.9,angle=0]{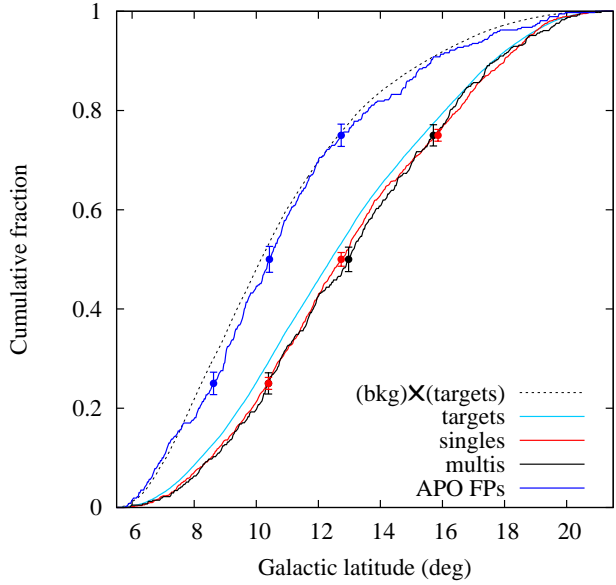}
\caption{
Distributions in galactic latitude of the 70,008 lowest noise \ik  targets that are classified as dwarf stars and were observed for at least five of the first eight quarters (solid turquoise curve), the 1311 targets with exactly one planet
candidates meeting the six criteria for our statistical study listed at the beginning of Section 3 (solid  red curve), and the 412 targets with multiple qualifying candidates (solid black curve).  The dashed  curve shows the
expected distribution of  chance alignment blends, and the purple curve represents the distribution of KOIs that are  labeled as FPs because their location on the sky plane differs from that of the target star (such FPs are known as active pixel offsets, APOs). 
}
\label{fig:gallat}
\end{figure}

Figure \ref{fig:gallat} shows as solid curves the cumulative distributions in galactic latitude of \ik targets, planet candidates and false positives that are spatially offset from \ik targets.
The dashed  curve in the figure shows the
distribution of expected chance alignment blends, which we compute by weighting each
target by the sky density of stars with magnitudes between 15.0 and
20.0 according to the Besancon model of the galaxy (Robin et al.~2003,
2004).  Each curve is normalized to give a total of unity.  Error bars represent the statistical uncertainty 
(1$\sigma$ binomial variation) of the three curves that represent populations small enough to have noticeable stochastic variations at quartile points in their 
distributions. 

The basic
shape of the target and planet candidate curves in Figure \ref{fig:gallat} is governed by the orientation of the
\ik field on the sky, where the amount of area covered increases from
a point at galactic latitude $5.5^\circ$ to a region spanning $16^\circ$ in
longitude at $13^\circ$ in galactic latitude and then decreases to a
point at $21.5^\circ$ latitude; the second most important factor is a
weighting towards the galactic plane because of the higher spatial
density of target stars in that portion of the field.  The purple and dashed 
curves rise rapidly at low  galactic latitudes because, to
first approximation, they are weighted quadratically in the spatial
density of stars (more precisely, the product of the spatial densities
of target stars and background stars), whereas the solid curves are
only weighted linearly.  The spatial density of background stars drops
by a factor of 15 from the lowest galactic latitude that \ik observes to the highest
latitude. 

The similarity of curves for targets, planets and multis in Figure \ref{fig:gallat},
and the difference of these from the curves showing estimated chance alignment blends and observed BGEB FPs, suggest
that the fraction of surviving candidates that are produced by BGEBs (as well as
that resulting from planets transiting background stars) is small. The combination of statistical uncertainties and unmodeled effects such as the dependance of aperture size on crowding in the star field and distance from the center of the \ik field would need to be analyzed in greater detail to determine whether or not there is any significance to the small differences between the curves for targets and planet candidates. 

\section{ Appendix B: Few Planets in Multis have Periods $< 1.6$ Days}

Planetary transits and stellar eclipses produce periodic dips in the light curves of thousands of \ik targets. Comparing the distributions of periods of various classes of events provides both astrophysical insight and information useful for planet validation.  The ensemble of planet candidates in multis (PCMs) has a smaller fraction of members with short periods than do the viable single planet candidates found by \ikt. Identified false positives are even more concentrated at shorter orbital periods than are singles, and detached eclipsing binaries (that were initially recognized as such and thus were not given KOI numbers prior to 2013) are the most skewed to shorter periods (Figure \ref{fig:perdist}). The curve representing the cumulative period distribution of PCMs is concave upwards below periods of  3 -- 4  days, then flattens out and gradually turns over at periods of 8 -- 9 days. Very few PCMs have periods  $P < 1.6$ days.   
We first consider the differences between the distributions of multis and those of FPs and EBs, concentrating on their implications for reliability of planets that we validate by multiplicity.  We then turn to differences between single and multiple planet candidates, which may provide clues to the structure of planetary systems. 

False positives (whose light curve signatures look very much like those of transiting planets) are thought to be primarily caused by eclipsing binaries, so why is the period distribution of FPs intermediate between those of EBs and transiting planets at short periods?  One possibility is that many of the FPs are actually transiting planets, albeit perhaps not planets transiting the \ik target star.  An alternative, more likely, explanation for at least most of this difference is that a larger fraction of very short period EBs are easier to recognize as such because of the presence of secondary eclipses, tidal distortions, or other effects that more clearly distinguish the light curves of EBs that were not classified as KOIs  (starting in 2013, some of these more obvious EBs were given KOI numbers, but these numbers are larger than those of the range of KOIs considered for our study) from those of transiting planets at short periods.  Thus, we expect that the period distribution of recognized FPs is a better proxy for the distribution of unrecognized FPs that pollute the sample of (single and multiple) planet candidates than is the distribution of EBs.  

Using  the period distribution of recognized false positives as a proxy for the distribution of unrecognized FPs in the sample of planet candidates, we find that the ratio of the expected abundance of unidentified FPs in multis to planets in multis is far larger at short orbital periods than at long ones. As only 1.6\% of the PCMs have $P < 1.6$ days, compared to 27\% of identified FPs, the likelihood that a PCM with $P < 1.6$ days is an unrecognized FP is more than an order of magnitude larger than that of a member of a multi with $P > 1.6$ days.   We thus do not consider planet candidates with $P < 1.6$ days to be good prospects for validation by multiplicity. This isn't to imply that most or even any short-period PCMs are not true planets, just that the probability of an FP slipping through our tests, which use a statistical approach that incorporates the relative frequencies of planets and of FPs, is too large to consider them validated planets without a more detailed study than we perform.

Several factors may contribute to the much larger abundance of single planet candidates relative to that of  PCMs at very short periods ($P < 1.6$ days).   It is possible that this apparent difference is spurious, either because the detection efficiency of sister planets of transiting short-period planets is very small or the bulk of the singles at these short periods are FPs, but neither of these observational biases are likely to be strong enough to explain the magnitude of the difference observed (\citealt{Fressin:2013} and \citealt{Santerne:2013}; cf.~\citealt{Beauge:2013}).  Thus, it appears that, like hot jupiters (\citealt{Latham:2011}, \citealt{Wright:2012}), very hot planets of all sizes are more common in single transiting planet systems than they are in multiple transiting planet systems.  

It may be that very close-in planets truly lack nearby planetary neighbors, as is the case for hot jupiters (\citealt{Steffen:2012},  \citealt{Wright:2012}), and/or that very close-in planets tend to have substantial orbital inclinations relative to more distant planets in their systems.  In either case, tidal forces could be the culprit. Neighboring planets might excite orbital eccentricities enough to lead to tidal decay of orbits into the star or to such short orbital periods that they are not efficienty identified by the \ik pipeline.
 These very hot planets could have been scattered inwards and tidally circularized, with the scattering randomizing orbital planes and removing planets in nearby orbits.  Or they could have had their orbital inclinations damped to near the equatorial planes of their star, leading to non-planar systems.  In this regard, note that the orbits of the two planets transiting Kepler-10, one of which has a period of 0.8 days, are inclined relative to one another by more than $5^\circ$ \citep{Batalha:2011, Fressin:2011}.  \cite{Steffen:2013b} have noted that few planets with $P < 3$ days have neighbors with small period ratios; this lack of close neighbors may be related to that of the overall deficit of planets with $P < 1.6$ days in multis.

There is also a small deficit of PCMs relative to singles at periods above 100 days.  Further study is required to determine whether or not this is the consequence of observational factors such as the difficulty in detecting small planets at long periods combined with the tendency for fewer large planets to be seen in multis and differences in the processes of detecting singles and multis.  If the difference reflects the true abundances of transiting planets, at least part of the cause is likely to be the smaller geometric probability of transits at longer periods yielding a smaller fraction of those multiple planet systems that lack close-in planets showing a second set of transits given one set being observed.

\section{ Appendix C: The Flip Side of the Reliability of Multis: Larger False Positive Probabilities for Single Planet Candidates}

Tools that aim to validate planet candidates statistically (e.g.,
BLENDER; \citealt{Torres:2011, Fressin:2011}, and the methods developed by \citealt{Morton:2012}) require as input the base
rates of the various astrophysical objects being considered, e.g.,
abundances of planets vs.~binary stars.
Previous studies have assumed base rates for planets (sometimes referred to as planet abundances or planet priors) derived
from the \ik planet candidates observed in systems with both single and
multiple planet candidates.
When attempting to validate targets with a single transiting planet
candidate, a more appropriate base rate would be derived from only the
single \ik planet candidates.  For the \cite{Fressin:2013} study, this planet occurrence rate is about 1/3 less than the values
computed using all \ik planet candidates, since roughly 1/3 of the \ik
planet candidates  in the
\cite{Batalha:2013} cumulative catalog that formed the basis of the
study by \cite{Fressin:2013} are members of multiple planet (candidate) systems.  We caution that this fraction depends
strongly on planet size and also somewhat on orbital period.
This implies that the value of  ${\cal P}_1$  (i.e., the probability
of a single transiting planet candidate being a planet) is smaller
than the estimated values quoted for the planet candidate population
as a whole by \cite{Fressin:2013}.  

As $1 - {\cal P}_1  \ll 1$, this correction increases the
value of $1 - {\cal P}_1$  by roughly 50\% if only eclipsing
binaries are considered FPs.
The required adjustment is smaller in magnitude when one considers
planets orbiting the \ik target relative to planets orbiting stars
other than the target. Nevertheless, an increase in $1 - {\cal P}_1$  would still be present because the planets would
need to be larger if they orbited a fainter star\footnote{A smaller
planet could produce the same amplitude drop in the light curve if it
orbited a star that was substantially hotter (higher surface
brightness) than the target star or was much closer to the telescope
than is the target star, but these situations are not likely to be
common.} and the fraction of large planets found in multiple planet
systems is smaller than that for small planets \citep{Latham:2011}.
Note also that the required adjustments are smaller in magnitude for
larger planet candidates than for smaller planet candidates because of
the greater concentration of small planets in multiple planet systems. As the fraction of planet candidates in multis is higher for  more recent catalogs such as the list used for our statistical study and the one presented in \cite{Burke:2013}, the magnitude of this correction for singles is also larger if the same methodology is used.

\section{ Appendix D: Dynamical Simulations of Systems with Large and/or Tightly-Packed Planet Candidates} 

Two of our criteria for planet validation require dynamical studies of systems under consideration.  Criterion ($ix$) implies that we need to remove systems that would be unstable for reasonable planetary masses.    Criterion ($x$) allows us to validate objects large enough to be brown dwarfs only if dynamical limits on their masses from upper limits to observed transit timing variations of neighboring planets require them to be of planetary mass.  We discuss the dynamical analyses we conduct in support of planet validation in this appendix.

We perform two tests for stability. Since stability depends on planetary masses and most of the planetary candidates do not have measured masses, for both of these tests we adopt a simple mass-radius relation employed by \cite{Lissauer:2011b} and \cite{Fabrycky:2013}:
\begin{eqnarray}
M_p = \left(\frac{R_{p}}{R_{\oplus}} \right)^{2.06}M_{\oplus} \; \; \;  \; \; \; {\rm for}  \; \; \;\; R_p > R_{\oplus} ,\\
M_p = \left(\frac{R_{p}}{R_{\oplus}} \right)^{3}M_{\oplus} \; \;\;  \; \; \; {\rm for} \; \; \; \; R_p < R_{\oplus} .
\label{mr-relation}
\end{eqnarray}

We use Hill's criterion to test for stability of neighboring pairs of planet candidates. The mutual Hill radius, $R_{H\;j,j+1}$, of two neighboring planets ($j$ and $j+1$) depends on their masses, $M_j$ and $M_{j+1}$, and their orbital semimajor axes, $a_j$ and $a_{j+1}$, as:
\begin{equation}
R_{H\;j,j+1} = \left[\frac{M_j+M_{j+1}}{3M_{\star}} \right]^{1/3} \frac{(a_j+a_{j+1})}{2}.
\label{HillRadius}
\end{equation}

\noindent Hill's stability criterion for coplanar circular orbits is:
\begin{equation}
\Delta_{j,j+1} \equiv \frac{a_{j+1}-a_j}{R_{H\;j,j+1}} > 2\sqrt{3} \approx 3.46.
\label{HillsCriterionforPairs}
\end{equation}
Two multis, KOI-284 and KOI-2248, fail this stability test by amounts that cannot be explained by any plausible errors in estimates of planetary masses, as noted by  \cite{Lissauer:2011b} and \cite{Fabrycky:2013}, respectively. These two multis are discussed in Section 6.2. Other systems that fail Hill's criterion have high estimated impact parameters, which disqualifies them for validation herein, and  they probably fail to satisfy Equation (\ref{HillsCriterionforPairs})  because their  radii and hence their masses are substantially overestimated.

Hill's stability criterion does not account for the additional destabilizing influence of multi-planet interactions present in systems with three or more planets. To find KOIs whose validity is dubious because of instabilities induced by the interactions of three or more planets, we   follow  \cite{Fabrycky:2013} and perform numerical integrations on systems where three planets are so close that
\begin{equation}
\Delta_{j-1,j}+\Delta_{j,j+1} < 18.
\label{HillsCriterionforThrees}
\end{equation}
The three multi-planet systems  KOI-157 (Kepler-11), KOI-707 (Kepler-33), KOI-806 (Kepler-30) have such  tightly packed planets, but have been verified by detailed studies and already been tested for dynamical stability.  Other candidates that Expression (\ref{HillsCriterionforThrees}) calls into question include KOI-1831 and KOI-1426, whose candidates have impact parameters so high as to preclude validation by our protocol  and thus are not considered further.  Four multis, KOI-152 (Kepler-79), KOI-620 (Kepler-51),  KOI-886 (Kepler-54), and KOI-1102 (Kepler-24), are tightly packed, meet our other validation criteria, and have candidates that were not previously confirmed as planets\footnote{A contemporaneous study by \cite{Jontof:2013} confirms all four of Kepler-79's candidates as planets and derives estimates of their masses and eccentricities}. We integrate these four systems, assuming initially circular and coplanar orbits and adopting the  mass-radius relation given by Equations (11) and (\ref{mr-relation}) for 1 Gyr to test for stability through direct simulations. Our integrations  used the HNBody symplectic integrator with 1/30th of the orbit period of the innermost candidate as our timestep \citep{Rouch:2002}. Each system remained stable for the entire 1 Gyr time interval simulated.

We use a different protocol to validate ``giant'' candidates whose  radii are consistent with those of brown dwarfs, i.e., $R_p + 2\sigma_{R_p} > 9.0~R_{\oplus}$. For these objects, we conducted dynamical fits to  transit times observed in Q1 -- Q14  \ik long cadence data, to determine whether the giant candidates can have masses exceeding 13 $M_{J}$ (Jupiter masses), assuming all of the candidates  orbit the \ik target. We integrate our dynamical models with an 8th order Dormand-Prince Runge-Kutta integrator to generate simulated transit times,  compare these times to our observed dataset and perform Levenberg-Marquart minimization to find the best fit. Our free parameters include the orbital periods at the epoch JD = 2,455,680 and the time of the first transit after this epoch. We assume coplanar orbits and fix the giant planet masses at $13~M_{J}$. We also adopt circular orbits at epoch throughout to be conservative in our validation protocol, since nonzero eccentricities typically produce a similar range of simulated TTVs for a lower mass.  We fix the masses of the remaining  candidates in each system at values determined by the same mass-radius relation adopted in our stability analysis. If the simulated TTVs clearly exceed the observed TTVs in the candidate or candidates neighboring the giant, the giant cannot be a brown dwarf or secondary star in the same system as the other candidate(s), so it passes criterion ($x$) and remains eligible for validation. If the simulated TTVs do not clearly exceed the observed TTVs, we perform another fit to the transit times, letting the mass of the giant float, and we measure the change in $\chi^2$ per degree of freedom (d.o.f.). If the measured mass $M_{p} < 13~M_{J}$ at the best fit solution, and if the lower mass solution improves on the fit   by   $\Delta \chi^2/(\rm{d.o.f.}) > 4$ relative to the fit with the giant's mass fixed at $ 13~M_{J}$, there is evidence at the  $95 \%$ confidence level that the candidate's mass is too low for it to be considered a brown dwarf or a star and we consider it to have passed criterion ($x$). Otherwise, the  candidate giant planet is left unvalidated. Note that none of the giant planet candidates had best-fit solutions with $M_{p} > 13~M_{J}$, so these unvalidated candidates may well be real planets.


\end{document}